\documentclass[english,12pt]{article}
\usepackage[T1]{fontenc}
\usepackage[latin1]{inputenc}
\usepackage{hyperref}

\usepackage{geometry}
\usepackage{epsfig}
\usepackage{cite}
\usepackage{color}
\usepackage{comment}
\usepackage{amsmath}
\usepackage{amssymb}
\usepackage{multirow}
\usepackage{bigdelim}
\usepackage{booktabs}
\usepackage{longtable}

\def\({\left(} 
\def\){\right)}

\newcommand{\VBFNLO}{\textsc{Vbfnlo}}
\newcommand{\rot}[1]{\rlap{\rotatebox{45}{#1}}\hspace*{1em}}

\makeatletter
\def\instring#1#2{TT\fi\begingroup
  \edef\x{\endgroup\noexpand\in@{#1}{#2}}\x\ifin@}
\makeatother
\newcommand{\makeflag}[3]{%
\if\instring{#1}{#3}{$\checkmark$}\else\if\instring{#2}{#3}{$\bigcirc$}\else{$-$}\fi\fi%
}
\newcommand{\bsmoptions}[1]{%
\makeflag{L}{l}{#1} % semi-leptonic decay
&
\makeflag{F}{f}{#1} % VBF process
&
\makeflag{V}{v}{#1} % anomalous Gauge couplings
&
\makeflag{H}{h}{#1} % anomalous Higgs couplings
&
\makeflag{T}{t}{#1} % Two-Higgs model
&
\makeflag{K}{k}{#1} % Kaluza-Klein model
&
\makeflag{S}{s}{#1} % Spin-2 model
&
\makeflag{M}{m}{#1} % MSSM
}
\newcommand{\bsmgfoptions}[1]{%
\makeflag{G}{g}{#1} % gluon-fusion process
&
\makeflag{L}{l}{#1} % semi-leptonic decay
&
\makeflag{H}{h}{#1} % anomalous Higgs couplings
&
\makeflag{T}{t}{#1} % general 2HDM
&
\makeflag{M}{m}{#1} % MSSM
}

\begin{document}
\begin{titlepage}
{\begin{flushright}
FTUV-14-2903, IFIC-14-26, KA-TP-10-2014 \\
LPN14-062, MAN/HEP/2014/03
\end{flushright}}
\vspace{0.8cm}
\begin{center} 
  {\Large \bf Release Note -- \VBFNLO~2.7.0} 
\end{center}
\vspace{0.4cm}
\begin{center}
{\renewcommand{\baselinestretch}{4}
J.~Baglio$^{1}$, J.~Bellm$^{1}$, F.~Campanario$^{2}$,
B.~Feigl$^{1}$, J.~Frank$^{1}$, T.~Figy$^{3}$, 
M.~Kerner$^{1}$, L.D.~Ninh$^{1,4}$, S.~Palmer$^{1}$, M.~Rauch$^{1}$,
R.~Roth$^{1}$, F.~Schissler$^{1}$, O.~Schlimpert$^{1}$, D.~Zeppenfeld$^{1}$
}
\end{center}
\vspace{0.3cm}
\begin{center}
$^{1}$ Institute~for~Theoretical~Physics, Karlsruhe~Institute~of~Technology~(KIT), 76128~Karlsruhe, Germany \\ \noindent
$^{2}$ Theory~Division, IFIC, University~of~Valencia-CSIC, E-46980~Paterna, Valencia, Spain \\ \noindent
$^{3}$ School~of~Physics~and~Astronomy, The~University~of~Manchester, Manchester, M13~9PL, United~Kingdom\\ \noindent
$^{4}$ Institute~of~Physics, Vietnam~Academy~of~Science~and~Technology, 10~Dao~Tan, Ba~Dinh, Hanoi, Vietnam \\ \noindent
\noindent
\noindent
\end{center}
\vspace{0.4cm}

\begin{abstract}

\noindent\VBFNLO{} is a flexible parton level Monte Carlo program for the
simulation of vector boson fusion (VBF), double and triple vector boson (plus
jet) production as well as QCD-induced single and double vector boson production
plus two jets in hadronic collisions at next-to-leading order~(NLO) in the
strong coupling constant. Furthermore, Higgs boson plus two jet production via
gluon fusion at the one-loop level is included. This note briefly describes the main
additional features and processes that have been added in the new release --
\VBFNLO\textsc{ Version 2.7.0}.  

At NLO QCD several new processes are available. These are $W^\pm\gamma$ and $HH$
production in VBF, the QCD-induced single and double vector boson plus
two jets processes $W^\pm jj$, $W^\pm Zjj$, $W^\pm\gamma jj$, same-sign
$W^\pm W^\pm jj$, and the Higgs-strahlung process $W^\pm H$ (plus jet) production.
Anomalous couplings are now available for all VBF, diboson and triboson
(plus jet) processes including the new VBF $W^\pm\gamma jj$ 
implementation, as well as in the Higgs-strahlung (plus jet) process.

Semi-leptonic decay modes are supported for the following diboson,
triboson and VBF processes: $VV$, $VVV$, $VV\gamma$ production and $VV$
production via VBF, where $V$ denotes a massive gauge boson, i.e.\
$W^\pm$ or $Z$. Additionally, the VBF-Higgs production process with
decay into $WW$ or $ZZ$ supports semi-leptonic decays of the gauge
bosons. All these semi-leptonic processes contain the possibility to
include anomalous gauge boson couplings. 

\end{abstract}
\vspace{0.3cm}
\today
\end{titlepage}

\section{\textsc{Introduction}}

\VBFNLO~\cite{Arnold:2008rz,Arnold:2011wj,Arnold:2011wjv2,Arnold:2012xn} is a
flexible Monte Carlo (MC) program for vector boson fusion (VBF), double
and triple vector boson (plus jet) as well as QCD-induced single and
double vector boson plus two jets production processes at NLO QCD
accuracy. The electroweak corrections to on-shell Higgs boson production via VBF
have been included. In addition, the simulation of $\cal{CP}$-even and
$\cal{CP}$-odd Higgs boson production in gluon fusion, associated with
two additional jets, is implemented at the (one-loop) leading-order (LO) QCD level.
\VBFNLO{} can be run in the Minimal Supersymmetric Standard Model (MSSM), and
anomalous couplings of the
Higgs boson and gauge bosons have been implemented for a large fraction
of the available processes.  Additionally, a model for spin-2
resonances, the option to include two Higgs resonances
and two Higgsless extra dimension models -- the Warped Higgsless
scenario and a Three-Site Higgsless Model -- are included for selected
processes.

Further information, and the latest version of the code, can be found on the
\VBFNLO{} webpage
\begin{center}
{\tt http://www.itp.kit.edu/vbfnlo/}\, .
\end{center}
A complete process list is given in Appendix A.

%--------------------------------------------------------------------------------
%================= Section ======================================================
%--------------------------------------------------------------------------------

\section{New processes}

For the latest version of \VBFNLO{} we have implemented several new processes
at NLO QCD.

\subsection{QCD-induced single and double vector boson production plus
two jets}

The QCD-induced production of $Vjj$ and $VVjj$ events at
$\mathcal{O}(\alpha_s^2)$ in the strong coupling constant is a new
process class available in \VBFNLO~2.7.0. With the same
final state particles as the corresponding VBF processes they are an
irreducible background to the latter. Hence, they play an important role when
studying anomalous quartic gauge couplings. They also form a relevant
background in searches for physics beyond the Standard Model.
In \VBFNLO~2.7.0, the production of $W^\pm jj$ as single boson final state and
the diboson final states $W^\pm Z jj$~\cite{Campanario:2013qba}, $W^\pm
\gamma jj$~\cite{Campanario:2014dpa} and same-sign $W^\pm W^\pm
jj$~\cite{Campanario:2013gea} have been included. They are available at
NLO QCD accuracy including leptonic decays of the vector bosons and full
off-shell and finite-width effects. Instabilities in the virtual
amplitudes, which contain diagrams up to hexagon loops, can be cured by
resorting to quadruple precision for problematic points, if supported by
the compiler\footnote{This has been tested with gfortran 4.7.3 and 4.8.2
as well as ifort 12.1.0.}. The impact of the NLO QCD corrections on the
total cross section is modest for a reasonable choice of factorization and
renormalization scales, but reduces the scale uncertainty
significantly. In kinematic distributions a sizable phase-space
dependence is observed, which leads to relevant changes in the shape of
distributions.
The corresponding process IDs are given in Tables~\ref{tab:qcdvjj}
and~\ref{tab:qcdvvjj}.

\begin{table}[!htb]
\begin{center}
\small
\begin{tabular}{c|l}
\hline
&\\
\textsc{ProcId} & \textsc{Process}\\
&\\
\hline
&\\
\bf 3130 & $p \overset{\mbox{\tiny{(--)}}}{p} \to W^{+} \,  jj\to \ell^{+} \nu_\ell \, jj$ \\
\bf 3140 & $p \overset{\mbox{\tiny{(--)}}}{p} \to W^{-} \, jj\to \ell^{-} \bar{\nu}_\ell  \, jj$\\
&\\
\hline
\end{tabular}
\caption {\em  Process IDs for QCD induced vector boson + 2 jet production 
at NLO QCD accuracy.}
\vspace{0.2cm}
\label{tab:qcdvjj}
\end{center}
\end{table}
\begin{table}[!htb]
\begin{center}
\small
\begin{tabular}{c|l}
\hline
&\\
\textsc{ProcId} & \textsc{Process}\\
&\\
\hline
&\\
\bf 3220 & $p \overset{\mbox{\tiny{(--)}}}{p} \to W^{+}Z \,  jj\to \ell_{1}^{+} \nu_{\ell_{1}} \ell_{2}^{+} \ell_{2}^{-} \, jj$  \\
\bf 3230 & $p \overset{\mbox{\tiny{(--)}}}{p} \to W^{-}Z \, jj\to \ell_{1}^{-} \bar{\nu}_{\ell _{1}} \ell_{2}^{+} \ell_{2}^{-} \, jj$  \\
\bf 3250 & $p \overset{\mbox{\tiny{(--)}}}{p} \to W^{+}W^{+} \,  jj\to \ell_{1}^{+} \nu_{\ell_{1}} \ell_{2}^{+} \nu_{\ell_{2}} \, jj$ \\
\bf 3260 & $p \overset{\mbox{\tiny{(--)}}}{p} \to W^{-}W^{-} \,  jj\to \ell_{1}^{-} \bar{\nu}_{\ell_{1}} \ell_{2}^{-} \bar{\nu}_{\ell_{2}} \, jj$ \\
\bf 3270 & $p \overset{\mbox{\tiny{(--)}}}{p} \to W^{+}\gamma \,  jj\to \ell^{+} \nu_{\ell} \gamma \, jj$  \\
\bf 3280 & $p \overset{\mbox{\tiny{(--)}}}{p} \to W^{-}\gamma \, jj\to \ell^{-} \bar{\nu}_{\ell} \gamma \, jj$  \\
&\\
\hline
\end{tabular}
\caption {\em  Process IDs for QCD induced diboson + 2 jet
production at NLO QCD accuracy.}
\vspace{0.2cm}
\label{tab:qcdvvjj}
\end{center}
\end{table}

\subsection{$W^\pm\gamma$ production via VBF}

The list of implemented VBF processes has been extended by the new
$W^\pm\gamma$ final state~\cite{kaiser,Campanario:2013eta} in
\VBFNLO~2.7.0. As with other VBF processes, it allows for testing the
structure of electro-weak quartic gauge couplings and appears as a
background in searches for physics beyond the Standard Model.
Scale uncertainties at NLO QCD are significantly reduced compared to the
LO prediction and $K$ factors are typically close to unity.
The process IDs for these processes can be found in Table~\ref{tab:VBFWA}.

\begin{table}[!htb]
\begin{center}
\small
\begin{tabular}{c|l}
\hline
&\\
\textsc{ProcId} & \textsc{Process} \\
&\\
\hline
&\\
\bf 270 & $p \overset{\mbox{\tiny{(--)}}}{p} \to W^{+}\gamma \, jj\to \ell^{+} \nu_{\ell} \gamma \, jj$ \\
\bf 280 & $p \overset{\mbox{\tiny{(--)}}}{p} \to W^{-}\gamma \, jj\to \ell^{-} \bar{\nu}_{\ell} \gamma \, jj$ \\
&\\
\hline
\end{tabular}
\caption {\em  Process IDs for $W^\pm\gamma$ + 2 jet
production via vector boson fusion at NLO QCD accuracy.}
\vspace{0.2cm}
\label{tab:VBFWA}
\end{center}
\end{table}

\subsection{Double Higgs production via VBF}

The production of two Higgs bosons in
VBF~\cite{Figy:2008zd,Baglio:2012np} contains Feynman diagrams with
triple Higgs couplings and therefore allows to study the Higgs
self-coupling. The NLO QCD corrections are modest, of the order of several
percent for the total cross section, and theory
uncertainties due to scale variation are reduced compared to LO. Decays of
the Higgs bosons are not implemented so far, and the Higgs bosons are
produced on-shell. The corresponding process ID is listed in
Table~\ref{tab:hhjj}.

\begin{table}[htb!]
\begin{center}
\small
\begin{tabular}{c|l}
\hline
&\\
\textsc{ProcId} & \textsc{Process}\\
&\\
\hline
&\\
\bf 160 & $p \overset{\mbox{\tiny{(--)}}}{p} \to HH \,  jj$ \\
&\\
\hline
\end{tabular}
\caption {\em  Process ID for Higgs pair + 2 jet
production via vector boson fusion at NLO QCD accuracy.}
\vspace{0.2cm}
\label{tab:hhjj}
\end{center}
\end{table}

%--------------------------------------------------------------------------------
%================= Section ======================================================
%--------------------------------------------------------------------------------

\subsection{$W$-Higgs associated production with up to one jet}

Associated production of a Higgs boson and a $W^\pm$, also known as
Higgsstrahlung, is one of the main Higgs production modes at the LHC. It
is particularly important when the Higgs boson decays into bottom
quarks, which has the largest branching ratio. There, boosted topologies
allow one to enhance this process over the background and deduce
information on the bottom Yukawa coupling.  In \VBFNLO~2.7.0, this
process is implemented at NLO QCD, both without and with an additional
hard jet in the final-state~\cite{robin}. The corrections to the integrated cross
section show $K$ factors of approximately 1.5 and 1.2, respectively, which
is a typical size for diboson processes, and lead to reduced scale
variation uncertainties.  As a cross-check, also the processes without a
Higgs boson in the final-state, i.e.\ $W^\pm$ and $W^\pm + \text{ jet}$
production at NLO QCD have been calculated and are made available in
this release.
The implementation of all these processes includes leptonic decays and
off-shell effects of the $W$ boson. The Higgs is produced on-shell, but
one can include its decays into a variety of final states. 
The new process IDs are given in Tables~\ref{tab:wj}, \ref{tab:wh}
and~\ref{tab:whj}.

\begin{table}[!htb]
\begin{center}
\small
\begin{tabular}{c|l}
\hline
&\\
\textsc{ProcId} & \textsc{Process}  \\
&\\
\hline
&\\
\bf 1330 & $p \overset{\mbox{\tiny{(--)}}}{p} \to W^+ \to \ell^+\nu_{\ell} $ \\
\bf 1340 & $p \overset{\mbox{\tiny{(--)}}}{p} \to W^- \to \ell^- \bar{\nu}_{\ell} $ \\
\bf 1630 & $p \overset{\mbox{\tiny{(--)}}}{p} \to W^+ \, j \to \ell^+\nu_{\ell} \, j $ \\
\bf 1640 & $p \overset{\mbox{\tiny{(--)}}}{p} \to W^- \, j\to \ell^- \bar{\nu}_{\ell} \, j $ \\
&\\
\hline
\end{tabular}
\caption {\em  Process IDs for the W production processes with up to one jet at NLO QCD accuracy.}
\vspace{0.2cm}
\label{tab:wj}
\end{center}
\end{table}

\begin{table}[!htb]
\begin{center}
\small
\begin{tabular}{c|l}
\hline
&\\
\textsc{ProcId} & \textsc{Process} \\
&\\
\hline
&\\
\bf 1300 & $p \overset{\mbox{\tiny{(--)}}}{p} \to W^+ H \to \ell^+\nu_{\ell} H $ \\
\bf 1301 & $p \overset{\mbox{\tiny{(--)}}}{p} \to W^+ H \to \ell^+\nu_{\ell} \gamma\gamma $ \\
\bf 1302 & $p \overset{\mbox{\tiny{(--)}}}{p} \to W^+ H \to \ell^+\nu_{\ell} \mu^+\mu^- $ \\
\bf 1303 & $p \overset{\mbox{\tiny{(--)}}}{p} \to W^+ H \to \ell^+\nu_{\ell} \tau^+\tau^- $ \\
\bf 1304 & $p \overset{\mbox{\tiny{(--)}}}{p} \to W^+ H \to \ell^+\nu_{\ell} b\bar{b} $ \\
\bf 1305 & $p \overset{\mbox{\tiny{(--)}}}{p} \to W^+ H \to W^+ W^{+}W^{-} \to \ell_{1}^+\nu_{\ell_{1}} \ell_{2}^+\nu_{\ell_{2}} \ell_{3}^- \bar{\nu}_{\ell_{3}}$ \\
\bf 1306 & $p \overset{\mbox{\tiny{(--)}}}{p} \to W^+ H \to W^+ ZZ \to \ell_{1}^+\nu_{\ell_{1}} \ell_{2}^+ \ell_{2}^- \ell_{3}^+ \ell_{3}^-$ \\
\bf 1307 & $p \overset{\mbox{\tiny{(--)}}}{p} \to W^+ H \to W^+ ZZ \to \ell_{1}^+\nu_{\ell_{1}} \ell_{2}^+ \ell_{2}^- \nu_{\ell_{3}}  \bar{\nu}_{\ell_{3}}$ \\
&\\
\hline
&\\
\bf 1310 & $p \overset{\mbox{\tiny{(--)}}}{p} \to W^- H \to \ell^- \bar{\nu}_{\ell} H $ \\
\bf 1311 & $p \overset{\mbox{\tiny{(--)}}}{p} \to W^- H \to \ell^- \bar{\nu}_{\ell} \gamma\gamma $ \\
\bf 1312 & $p \overset{\mbox{\tiny{(--)}}}{p} \to W^- H \to \ell^- \bar{\nu}_{\ell} \mu^+\mu^- $ \\
\bf 1313 & $p \overset{\mbox{\tiny{(--)}}}{p} \to W^- H \to \ell^- \bar{\nu}_{\ell} \tau^+\tau^- $ \\
\bf 1314 & $p \overset{\mbox{\tiny{(--)}}}{p} \to W^- H \to \ell^- \bar{\nu}_{\ell} b\bar{b} $ \\
\bf 1315 & $p \overset{\mbox{\tiny{(--)}}}{p} \to W^- H \to W^- W^{+}W^{-} \to \ell_{1}^-\bar{\nu}_{\ell_{1}} \ell_{2}^+\nu_{\ell_{2}} \ell_{3}^- \bar{\nu}_{\ell_{3}}$ \\
\bf 1316 & $p \overset{\mbox{\tiny{(--)}}}{p} \to W^- H \to W^- ZZ \to \ell_{1}^-\bar{\nu}_{\ell_{1}} \ell_{2}^+ \ell_{2}^- \ell_{3}^+ \ell_{3}^-$ \\
\bf 1317 & $p \overset{\mbox{\tiny{(--)}}}{p} \to W^- H \to W^- ZZ \to \ell_{1}^-\bar{\nu}_{\ell_{1}} \ell_{2}^+ \ell_{2}^- \nu_{\ell_{3}}  \bar{\nu}_{\ell_{3}}$ \\
&\\
\hline
\end{tabular}
\caption {\em  Process IDs for the WH production processes at NLO QCD accuracy.}
\vspace{0.2cm}
\label{tab:wh}
\end{center}
\end{table}

\begin{table}[tbh!]
\begin{center}
\small
\begin{tabular}{c|l}
\hline
&\\
\textsc{ProcId} & \textsc{Process} \\
&\\
\hline
&\\
\bf 1600 & $p \overset{\mbox{\tiny{(--)}}}{p} \to W^+ H \, j \to \ell^+\nu_{\ell} H \, j $ \\
\bf 1601 & $p \overset{\mbox{\tiny{(--)}}}{p} \to W^+ H \, j \to \ell^+\nu_{\ell} \gamma\gamma \, j $ \\
\bf 1602 & $p \overset{\mbox{\tiny{(--)}}}{p} \to W^+ H \, j \to \ell^+\nu_{\ell} \mu^+\mu^- \, j $ \\
\bf 1603 & $p \overset{\mbox{\tiny{(--)}}}{p} \to W^+ H \, j \to \ell^+\nu_{\ell} \tau^+\tau^- \, j $ \\
\bf 1604 & $p \overset{\mbox{\tiny{(--)}}}{p} \to W^+ H \, j \to \ell^+\nu_{\ell} b\bar{b} \, j $ \\
\bf 1605 & $p \overset{\mbox{\tiny{(--)}}}{p} \to W^+ H \, j \to W^+ W^{+}W^{-} \, j \to \ell_{1}^+\nu_{\ell_{1}} \ell_{2}^+\nu_{\ell_{2}} \ell_{3}^- \bar{\nu}_{\ell_{3}}\, j $ \\
\bf 1606 & $p \overset{\mbox{\tiny{(--)}}}{p} \to W^+ H \, j \to W^+ ZZ \, j \to \ell_{1}^+\nu_{\ell_{1}} \ell_{2}^+ \ell_{2}^- \ell_{3}^+ \ell_{3}^-\, j $ \\
\bf 1607 & $p \overset{\mbox{\tiny{(--)}}}{p} \to W^+ H \, j \to W^+ ZZ \, j \to \ell_{1}^+\nu_{\ell_{1}} \ell_{2}^+ \ell_{2}^- \nu_{\ell_{3}}  \bar{\nu}_{\ell_{3}}\, j $ \\
&\\
\hline
&\\
\bf 1610 & $p \overset{\mbox{\tiny{(--)}}}{p} \to W^- H \, j \to \ell^- \bar{\nu}_{\ell} H \, j $ \\
\bf 1611 & $p \overset{\mbox{\tiny{(--)}}}{p} \to W^- H \, j \to \ell^- \bar{\nu}_{\ell} \gamma\gamma \, j $ \\
\bf 1612 & $p \overset{\mbox{\tiny{(--)}}}{p} \to W^- H \, j \to \ell^- \bar{\nu}_{\ell} \mu^+\mu^- \, j $ \\
\bf 1613 & $p \overset{\mbox{\tiny{(--)}}}{p} \to W^- H \, j \to \ell^- \bar{\nu}_{\ell} \tau^+\tau^- \, j $ \\
\bf 1614 & $p \overset{\mbox{\tiny{(--)}}}{p} \to W^- H \, j \to \ell^- \bar{\nu}_{\ell} b\bar{b} \, j $ \\
\bf 1615 & $p \overset{\mbox{\tiny{(--)}}}{p} \to W^- H \, j \to W^- W^{+}W^{-} \, j \to \ell_{1}^-\bar{\nu}_{\ell_{1}} \ell_{2}^+\nu_{\ell_{2}} \ell_{3}^- \bar{\nu}_{\ell_{3}}\, j $ \\
\bf 1616 & $p \overset{\mbox{\tiny{(--)}}}{p} \to W^- H \, j \to W^- ZZ \, j \to \ell_{1}^-\bar{\nu}_{\ell_{1}} \ell_{2}^+ \ell_{2}^- \ell_{3}^+ \ell_{3}^-\, j $ \\
\bf 1617 & $p \overset{\mbox{\tiny{(--)}}}{p} \to W^- H \, j \to W^- ZZ \, j \to \ell_{1}^-\bar{\nu}_{\ell_{1}} \ell_{2}^+ \ell_{2}^- \nu_{\ell_{3}}  \bar{\nu}_{\ell_{3}}\, j $ \\
&\\
\hline
\end{tabular}
\caption {\em  Process IDs for the WH plus jet production processes at NLO QCD accuracy.}
\vspace{0.2cm}
\label{tab:whj}
\end{center}
\end{table}

\section{New and extended features}

In addition to the new processes described above, several existing
calculations have been extended and new features added.

\subsection{Semi-leptonic decays}

Besides the fully leptonic decays, which are implemented by default in
\VBFNLO{}, vector bosons can also decay into a quark--anti-quark pair.
For several processes \VBFNLO{} has been extended to also include
semi-leptonic final states, i.e.\ final states with one vector boson decaying into a
quark--anti-quark pair and the other(s) into leptons~\cite{semilep}. As in the fully
leptonic decay modes, finite-width effects of the vector bosons as well
as off-shell and non-resonant contributions are included. These
processes have several types of applications. In experimental analyses
of the VBF processes, when looking for anomalous quartic gauge
couplings for example, both leptonic and semi-leptonic decay modes are
studied and the new implementation allows to get precise predictions for
these processes. Also, the semi-leptonic decay modes of the triboson
processes form the $s$-channel part of the corresponding VBF processes. When
studying phase-space regions where the VBF approximation is not
completely valid anymore, in particular for small invariant masses of
the two jets of the order of the $W$ and $Z$ mass, these can give
relevant contributions. The new implementation now allows to study their
effects.
Semi-leptonic decays for the vector bosons have been added for the
diboson processes with two massive vector bosons ($W^\pm W^\mp$, $W^\pm
Z$, $ZZ$), the VBF diboson processes with two massive vector bosons
($W^\pm W^\mp jj$, $W^\pm W^\pm jj$, $W^\pm Z jj$, $ZZ jj$), VBF Higgs
boson production with decays into $W^\pm W^\mp$ and $ZZ$ and all triple
vector boson production processes with zero or one final state photon.
NLO QCD corrections to the production part are fully included.
NLO QCD effects in the hadronic vector boson decay can be estimated in all cases
by switching on multiplication with the corresponding overall $K$ factor for
$V\rightarrow q \bar{q}$.
A complete list of process IDs, where semi-leptonic decays are
implemented, is given in Tables~\ref{tab:semilep1}
and~\ref{tab:semilep2}.
The flavour of the final-state quarks can be chosen by setting the
variable \texttt{DECAY\_QUARKS} in \texttt{vbfnlo.dat} to the
corresponding PDG IDs.

\begin{table}[htb!]
\begin{center}
\small
\begin{tabular}{c|l}
\hline
&\\
\textsc{ProcId} & \textsc{Process} \\
&\\
\hline
&\\
\bf 108 & $p \overset{\mbox{\tiny{(--)}}}{p} \to H \, jj\to W^{+}W^{-} \, jj\to q\bar{q} \, \ell^- 
\bar{\nu}_{\ell} \,jj$  \\
\bf 109 & $p \overset{\mbox{\tiny{(--)}}}{p} \to H \, jj\to W^{+}W^{-} \, jj\to \ell^+\nu_{\ell} \,
q\bar{q}  \,jj$  \\
\bf 1010 & $p \overset{\mbox{\tiny{(--)}}}{p} \to H \, jj\to ZZ \, jj\to q\bar{q} \, \ell^+ \ell^- \,jj$ \\
&\\
\hline
&\\
\bf 201 & $p \overset{\mbox{\tiny{(--)}}}{p} \to W^{+}W^{-} \, jj \to q \bar{q} \, \ell^{-}
\bar{\nu}_{\ell} \, jj$ \\
\bf 202 & $p \overset{\mbox{\tiny{(--)}}}{p} \to W^{+}W^{-}  \, jj\to \ell^{+} \nu_\ell \, q \bar{q} \, jj$ \\
\bf 212 & $p \overset{\mbox{\tiny{(--)}}}{p} \to ZZ  \, jj\to q \bar{q} \, \ell^{+} \ell^{-} \, jj$ \\
\bf 221 & $p \overset{\mbox{\tiny{(--)}}}{p} \to W^{+}Z \,  jj\to q \bar{q} \, \ell^{+} \ell^{-} \, jj$  \\
\bf 222 & $p \overset{\mbox{\tiny{(--)}}}{p} \to W^{+}Z \,  jj\to \ell^{+} \nu_{\ell} \, q \bar{q} \, jj$  \\
\bf 231 & $p \overset{\mbox{\tiny{(--)}}}{p} \to W^{-}Z \, jj\to q \bar{q} \, \ell^{+} \ell^{-} \, jj$  \\
\bf 232 & $p \overset{\mbox{\tiny{(--)}}}{p} \to W^{-}Z \, jj\to \ell^{-} \bar{\nu}_{\ell} \, q \bar{q} \, jj$  \\
\bf 251 & $p \overset{\mbox{\tiny{(--)}}}{p} \to W^{+}W^{+} \,  jj\to q \bar{q} \, \ell^{+} \nu_{\ell} \, jj$  \\
\bf 261 & $p \overset{\mbox{\tiny{(--)}}}{p} \to W^{-}W^{-} \,  jj\to q \bar{q} \, \ell^{-} \bar{\nu}_{\ell} \, jj$  \\
&\\
\hline
\end{tabular}
\caption {\em  Process IDs for the VBF production processes at NLO QCD accuracy with semileptonic decays.}
\vspace{0.2cm}
\label{tab:semilep1}
\end{center}
\end{table}

\begin{table}[t!]
\begin{center}
\small
\begin{tabular}{c|l}
\hline
&\\
\textsc{ProcId} & \textsc{Process} \\
&\\
\hline
&\\
\bf 301 & $p \overset{\mbox{\tiny{(--)}}}{p} \to W^{+}W^{-} \to q \bar{q} \, \ell^{-}\bar{\nu}_{\ell} $ \\
\bf 302 & $p \overset{\mbox{\tiny{(--)}}}{p} \to W^{+}W^{-} \to \ell^{+} \nu_{\ell} \, q \bar{q} $ \\
\bf 312 & $p \overset{\mbox{\tiny{(--)}}}{p} \to W^{+}Z \to  q \bar{q} \, \ell^{+} \ell^{-} $ \\
\bf 313 & $p \overset{\mbox{\tiny{(--)}}}{p} \to W^{+}Z \to  \ell^{+} \nu_{\ell} \, q \bar{q} $ \\
\bf 322 & $p \overset{\mbox{\tiny{(--)}}}{p} \to W^{-}Z \to q \bar{q} \, \ell^{+} \ell^{-} $  \\
\bf 323 & $p \overset{\mbox{\tiny{(--)}}}{p} \to W^{-}Z \to \ell^{-} \bar{\nu}_{\ell} \, q \bar{q} $  \\
\bf 331 & $p \overset{\mbox{\tiny{(--)}}}{p} \to ZZ \to q \bar{q} \, \ell^{-} \ell^{+} $ \\
&\\
\hline
&\\
\bf 401 & $p \overset{\mbox{\tiny{(--)}}}{p} \to W^{+}W^{-}Z \to q \bar{q} \, \ell_{1}^{-} \bar{\nu}_{\ell_{1}} 
\ell_{2}^{+} \ell_{2}^{-} $ \\
\bf 402 & $p \overset{\mbox{\tiny{(--)}}}{p} \to W^{+}W^{-}Z \to \ell_{1}^{+}\nu_{\ell_{1}} \, q \bar{q} \,
\ell_{2}^{+} \ell_{2}^{-} $ \\
\bf 403 & $p \overset{\mbox{\tiny{(--)}}}{p} \to W^{+}W^{-}Z \to \ell_{1}^{+}\nu_{\ell_{1}} \ell_{2}^{-} \bar{\nu}_{\ell_{2}} \,
q \bar{q} $ \\
\bf 411 & $p \overset{\mbox{\tiny{(--)}}}{p} \to ZZW^{+} \to  \ell_{1}^{+} \ell_{1}^{-}  \ell_{2}^{+} \ell_{2}^{-} 
 \, q \bar{q} $ \\
\bf 412 & $p \overset{\mbox{\tiny{(--)}}}{p} \to ZZW^{+} \to  q \bar{q} \, \ell_{1}^{+} \ell_{1}^{-}
 \ell_{2}^{+} \nu_{\ell_{2}} $ \\
\bf 421 & $p \overset{\mbox{\tiny{(--)}}}{p} \to ZZW^{-} \to \ell_{1}^{+} \ell_{1}^{-}  \ell_{2}^{+} \ell_{2}^{-} 
 \, q \bar{q} $ \\
\bf 422 & $p \overset{\mbox{\tiny{(--)}}}{p} \to ZZW^{-} \to q \bar{q} \, \ell_{1}^{+} \ell_{1}^{-}
 \ell_{2}^{-}  \bar{\nu}_{\ell_{2}}$ \\
\bf 431 & $p \overset{\mbox{\tiny{(--)}}}{p} \to W^{+}W^{-}W^{+} \to  q \bar{q} \, \ell_{1}^{-}
\bar{\nu}_{\ell_1} \ell_{2}^{+}\nu_{\ell_{2}}$ \\
\bf 432 & $p \overset{\mbox{\tiny{(--)}}}{p} \to W^{+}W^{-}W^{+} \to \ell_{1}^{+}\nu_{\ell_1} \, q \bar{q} \,
 \ell_{2}^{+}\nu_{\ell_{2}}$ \\
\bf 441 & $p \overset{\mbox{\tiny{(--)}}}{p} \to W^{-}W^{+}W^{-} \to \ell_{1}^{-} \bar{\nu}_{\ell_1} \, q \bar{q} \,
\ell_{2}^{-} \bar{\nu}_{\ell_{2}} $ \\
\bf 442 & $p \overset{\mbox{\tiny{(--)}}}{p} \to W^{-}W^{+}W^{-} \to  q \bar{q} \, \ell_{1}^{+}\nu_{\ell_1}
\ell_{2}^{-} \bar{\nu}_{\ell_{2}} $ \\
\bf 451 & $p \overset{\mbox{\tiny{(--)}}}{p} \to ZZZ \to  q \bar{q} \, \ell_{1}^{-} \ell_{1}^{+} \ell_{2}^{-}
\ell_{2}^{+}  $ \\
\bf 461 & $p \overset{\mbox{\tiny{(--)}}}{p} \to W^{+}W^{-} \gamma \to  q \bar{q} \,
\ell^{-}\bar{\nu}_{\ell} \gamma$ \\
\bf 462 & $p \overset{\mbox{\tiny{(--)}}}{p} \to W^{+}W^{-} \gamma \to \ell^{+} \nu_{\ell}
\, q \bar{q} \, \gamma$ \\
\bf 471 & $p \overset{\mbox{\tiny{(--)}}}{p} \to Z Z \gamma \to \ell^{-} \ell^{+} \, q \bar{q} \, \gamma$ \\
\bf 481 & $p \overset{\mbox{\tiny{(--)}}}{p} \to W^{+} Z \gamma \to  q \bar{q} \, \ell^{-}
\ell^{+} \gamma$ \\
\bf 482 & $p \overset{\mbox{\tiny{(--)}}}{p} \to W^{+} Z \gamma \to \ell^{+}\nu_{\ell} \, q \bar{q} \, \gamma$ \\
\bf 491 & $p \overset{\mbox{\tiny{(--)}}}{p} \to W^{-} Z \gamma \to  q \bar{q} \, \ell^{-}
\ell^{+} \gamma$ \\
\bf 492 & $p \overset{\mbox{\tiny{(--)}}}{p} \to W^{-} Z \gamma \to \ell^{-} \bar{\nu}_{\ell} \, q \bar{q} \, \gamma$ \\
&\\
\hline
\end{tabular}
\caption {\em  Process IDs for the diboson and triboson production processes at NLO QCD accuracy with semileptonic decays.}
\vspace{0.2cm}
\label{tab:semilep2}
\end{center}
\end{table}

Several issues arise with respect to the application of cuts for semi-leptonic
decays. The first one concerns the definition of the two tagging jets.
Choosing the two jets with the largest transverse momentum might not be
the most advantageous choice, as this will occasionally select a
vector boson decay product as tagging jet, which alters the distinct
shape of the tagging jet distributions. Therefore, several new
definitions and further options to define the tagging jets have been
added. 
Also, the two partons which are decay products of the vector boson might
be grouped into a single jet by the jet algorithm, or one of them might
get combined with one of the tagging jets in VBF processes. In both
cases there is no QCD singularity associated when these two partons
become collinear. Hence, a flag has been implemented to select whether
such configurations are allowed or not. 
There is, however, a problem with contributions from virtual photons,
which are part of the processes with hadronically decaying $Z$ bosons. If
their invariant mass approaches zero, a QED singularity arises. This can
happen either when the two partons are allowed to form a single jet, or
in the real emission process, when the actual extra emission is
identified as a separate jet, again allowing the two decay products to
become collinear. To handle these cases, a cut on the minimal
invariant mass of the vector boson can be placed. Alternatively, a cut
on $m_{q\bar{q}}$ is
estimated for each final-state quark flavour such that the NLO
approximation for $\sigma(e^+e^- \rightarrow \text{hadrons})$ gives the
same contribution as the experimental continuum data plus the contribution from
the sharp resonances of the respective quark flavours
\cite{Beringer:1900zz}.
This procedure approximates the correct rates from low-$q^2$ photons.
The kinematics of the quarks in this region are not modeled correctly,
but this is of minor importance as the low-$q^2$ region is only relevant
for semileptonic decays when the decay products form a single jet.
For a more detailed discussion of the various options and their
default values we refer the reader to the manual of
\VBFNLO-2.7.0\footnote{\href{http://www.itp.kit.edu/~vbfnloweb/wiki/doku.php?id=documentation:manual}{http://www.itp.kit.edu/$\sim$vbfnloweb/wiki/doku.php?id=documentation:manual}}.

\subsection{Anomalous couplings}
Anomalous triple and quartic gauge boson couplings have been implemented for 
the remaining processes of double vector boson production via VBF, $W^\pm Z jj$,
$ZZjj$, $W^\pm W^\pm jj$ and the newly added $W^\pm \gamma jj$.
Also in the triboson processes, anomalous couplings are now available
for all final-state configurations, namely 
the $ZZZ$ process, which was the only process with three massive bosons
still missing, and in all with one or more final-state photons,
including the triboson+jet process $W^\pm \gamma\gamma j$.
Additionally, for the triboson processes $W^\pm W^\mp W^\pm$, $W^\pm
W^\mp Z$ and $W^\pm ZZ$ the set of operators leading to anomalous triple
gauge boson couplings has been extended.
The gluon-fusion production of a Higgs plus two jets with decays into
$WW$ or $ZZ$ can now be used with anomalous Higgs couplings.
To run \textsc{Vbfnlo} with anomalous couplings, the switch {\tt
ANOM\_CPL} in the input file {\tt vbfnlo.dat} must be switched to {\tt
true}.  The anomalous coupling parameters are then input via {\tt
anomV.dat} or (for $HVV$ couplings) {\tt anom\_HVV.dat}. The process IDs
for the processes where anomalous couplings have been newly implemented or
extended are collected in Table~\ref{tab:anom}.

\begin{table}[htb!]
\begin{center}
\small
\begin{tabular}{c|l}
\hline
&\\
\textsc{ProcId} & \textsc{Process} \\
&\\
\hline
&\\
\bf 210 & $p \overset{\mbox{\tiny{(--)}}}{p} \to ZZ  \, jj\to \ell_{1}^{+} \ell_{1}^{-} \ell_{2}^{+} \ell_{2}^{-} \, jj$ \\
\bf 211 & $p \overset{\mbox{\tiny{(--)}}}{p} \to ZZ  \, jj\to \ell_{1}^{+} \ell_{1}^{-} \nu_{\ell_{2}} \bar{\nu}_{\ell_{2}} \, jj$  \\
\bf 220 & $p \overset{\mbox{\tiny{(--)}}}{p} \to W^{+}Z \,  jj\to \ell_{1}^{+} \nu_{\ell_{1}} \ell_{2}^{+} \ell_{2}^{-} \, jj$  \\
\bf 230 & $p \overset{\mbox{\tiny{(--)}}}{p} \to W^{-}Z \, jj\to \ell_{1}^{-} \bar{\nu}_{\ell _{1}} \ell_{2}^{+} \ell_{2}^{-} \, jj$  \\
\bf 250 & $p \overset{\mbox{\tiny{(--)}}}{p} \to W^{+}W^{+} \,  jj\to \ell_{1}^{+} \nu_{\ell_{1}} \ell_{2}^{+} \nu_{\ell_{2}} \, jj$ \\
\bf 260 & $p \overset{\mbox{\tiny{(--)}}}{p} \to W^{-}W^{-} \,  jj\to \ell_{1}^{-} \bar{\nu}_{\ell_{1}} \ell_{2}^{-} \bar{\nu}_{\ell_{2}} \, jj$  \\
\bf 270 & $p \overset{\mbox{\tiny{(--)}}}{p} \to W^{+}\gamma \, jj\to \ell^{+} \nu_{\ell} \gamma \, jj$ \\
\bf 280 & $p \overset{\mbox{\tiny{(--)}}}{p} \to W^{-}\gamma \, jj\to \ell^{-} \bar{\nu}_{\ell} \gamma \, jj$ \\
&\\
\hline
&\\
\bf 400 & $p \overset{\mbox{\tiny{(--)}}}{p} \to W^{+}W^{-}Z \to \ell_{1}^{+}\nu_{\ell_{1}} \ell_{2}^{-} \bar{\nu}_{\ell_{2}} 
\ell_{3}^{+} \ell_{3}^{-} $ \\
\bf 410 & $p \overset{\mbox{\tiny{(--)}}}{p} \to ZZW^{+} \to  \ell_{1}^{+} \ell_{1}^{-}  \ell_{2}^{+} \ell_{2}^{-} 
 \ell_{3}^{+} \nu_{\ell_{3}} $  \\
\bf 420 & $p \overset{\mbox{\tiny{(--)}}}{p} \to ZZW^{-} \to \ell_{1}^{+} \ell_{1}^{-}  \ell_{2}^{+} \ell_{2}^{-} 
 \ell_{3}^{-}  \bar{\nu}_{\ell_{3}}$  \\
\bf 430 & $p \overset{\mbox{\tiny{(--)}}}{p} \to W^{+}W^{-}W^{+} \to \ell_{1}^{+}\nu_{\ell_1} \ell_{2}^{-}
\bar{\nu}_{\ell_2} \ell_{3}^{+}\nu_{\ell_{3}}$  \\
\bf 440 & $p \overset{\mbox{\tiny{(--)}}}{p} \to W^{-}W^{+}W^{-} \to \ell_{1}^{-} \bar{\nu}_{\ell_1}\ell_{2}^{+}\nu_{\ell_2}
\ell_{3}^{-} \bar{\nu}_{\ell_{3}} $  \\
\bf 450 & $p \overset{\mbox{\tiny{(--)}}}{p} \to ZZZ \to \ell_{1}^{-} \ell_{1}^{+} \ell_{2}^{-}
\ell_{2}^{+} \ell_{3}^{-} \ell_{3}^{+} $ \\
\bf 460 & $p \overset{\mbox{\tiny{(--)}}}{p} \to W^{-}W^{+} \gamma \to \ell_{1}^{-} \bar{\nu}_{\ell_1}
\ell_{2}^{+}\nu_{\ell_2} \gamma$  \\
\bf 470 & $p \overset{\mbox{\tiny{(--)}}}{p} \to Z Z \gamma \to \ell_{1}^{-} \ell_{1}^{+} \ell_{2}^{-}
\ell_{2}^{+} \gamma$  \\
\bf 480 & $p \overset{\mbox{\tiny{(--)}}}{p} \to W^{+} Z \gamma \to \ell_{1}^{+}\nu_{\ell_1} \ell_{2}^{-}
\ell_{2}^{+} \gamma$  \\
\bf 490 & $p \overset{\mbox{\tiny{(--)}}}{p} \to W^{-} Z \gamma \to \ell_{1}^{-} \bar{\nu}_{\ell_1} \ell_{2}^{-}
\ell_{2}^{+} \gamma$  \\
\bf 500 & $p \overset{\mbox{\tiny{(--)}}}{p} \to W^{+} \gamma \gamma \to {\ell}^{+}\nu_{\ell} 
\gamma \gamma$  \\
\bf 510 & $p \overset{\mbox{\tiny{(--)}}}{p} \to W^{-} \gamma \gamma \to {\ell}^{-} \bar{\nu}_{\ell} 
\gamma \gamma$  \\
\bf 520 & $p \overset{\mbox{\tiny{(--)}}}{p} \to Z \gamma \gamma \to {\ell}^{-} {\ell}^{+} 
\gamma \gamma$  \\
\bf 521 & $p \overset{\mbox{\tiny{(--)}}}{p} \to Z \gamma \gamma \to \nu_{\ell} \bar{\nu}_{\ell} 
\gamma \gamma$  \\
\bf 530 & $p \overset{\mbox{\tiny{(--)}}}{p} \to \gamma \gamma \gamma $ \\
&\\
\hline
&\\
\bf 4105 & $p \overset{\mbox{\tiny{(--)}}}{p} \to H \, jj\to W^{+}W^{-} \, jj\to \ell_{1}^+\nu_{\ell_{1}} \ell_{2}^- 
\bar{\nu}_{\ell_{2}} \,jj$ \\
\bf 4106 & $p \overset{\mbox{\tiny{(--)}}}{p} \to H \, jj\to ZZ \, jj\to \ell_{1}^+ \ell_{1}^- \ell_{2}^+ 
\ell_{2}^- \,jj$  \\
\bf 4107 & $p \overset{\mbox{\tiny{(--)}}}{p} \to H \, jj\to ZZ \, jj\to \ell_{1}^+ \ell_{1}^- \nu_{\ell_{2}}  
\bar{\nu}_{\ell_{2}} \,jj$  \\
&\\
\hline
\end{tabular}
\caption {\em  Process IDs for existing processes which have been extended to include anomalous couplings.}
\vspace{0.2cm}
\label{tab:anom}
\end{center}
\end{table}

\subsection{Two-Higgs model}

For the diboson VBF processes with two massive vector bosons (200-260), which
contain a Higgs propagator, a model with two CP-even
Higgs bosons has been added. This can be used for example when studying
heavy Higgs resonances, to account for both the discovered resonance at
around 126 GeV and an additional heavy resonance. Setting the couplings
appropriately, the high-energy behaviour can be adjusted so that no
unitary violation occurs.
This model is chosen by setting \texttt{MODEL = 3}. The mass and width
of the first Higgs boson are given by \texttt{HMASS} and
\texttt{HWIDTH}, while those of the second one are steered by
\texttt{H2MASS} and \texttt{H2WIDTH}, respectively. The squared coupling
of the Higgs to electro-weak gauge bosons can be altered by a
multiplicative factor, namely \texttt{SIN2BA} for the first one and
\texttt{COS2BA} for the second one. The unitarity requirement mentioned
above is fulfilled by setting $\texttt{SIN2BA}+\texttt{COS2BA}=1$.
All variables can be found in \texttt{vbfnlo.dat}.

\subsection{Spin-2 model}

The list of processes available within the spin-2 model, 
introduced in Refs.~\cite{Frank:2012wh,Frank:2013gca}, has been
extended. The model uses an effective Lagrangian to
describe the interactions of spin-2 particles with electroweak gauge
bosons. Both an isospin singlet spin-2 state and a spin-2 triplet are
available. In the previous release the possibility of an additional
spin-2 resonance besides Higgs and continuum diagrams has been added to
the diboson VBF processes 200-230, and a spin-2 resonance decaying into
a pair of photons is available with process ID 191. With \VBFNLO-2.7.0,
also a single spin-2 resonance decaying into $WW\rightarrow 2\ell 2\nu$,
$ZZ\rightarrow 4\ell$ or $ZZ\rightarrow 2\ell2\nu$ is added. This
complements the existing full processes by allowing to study the signal
process separately on its own. Note that the switch \texttt{SPIN2} in
\texttt{vbfnlo.dat} needs to be set to \texttt{true} for these
processes. If this switch is set to \texttt{false}, the same process
but with a Higgs resonance is calculated, allowing for an easy
comparison between the two possibilities. The process IDs of the new
processes are listed in Table~\ref{tab:spin2}.

\begin{table}[t!]
\begin{center}
\small
\begin{tabular}{c|l}
\hline
&\\
\textsc{ProcId} & \textsc{Process} \\
&\\
\hline
&\\
\bf 195 & $p \overset{\mbox{\tiny{(--)}}}{p} \to S_{2}  \, jj\to W^{+}W^{-} \, jj\to \ell_{1}^+\nu_{\ell_{1}} \ell_{2}^- 
\bar{\nu}_{\ell_{2}} \,jj$ 
\\
\bf 196 & $p \overset{\mbox{\tiny{(--)}}}{p} \to S_{2}  \, jj\to ZZ \, jj\to \ell_{1}^+ \ell_{1}^- \ell_{2}^+ 
\ell_{2}^- \,jj$ 
\\
\bf 197 & $p \overset{\mbox{\tiny{(--)}}}{p} \to S_{2}  \, jj\to  ZZ \, jj\to \ell_{1}^+ \ell_{1}^- \nu_{\ell_{2}}  
\bar{\nu}_{\ell_{2}} \,jj$
\\
&\\
\hline
\end{tabular}
\caption {\em  Process IDs for a spin-2 particle $S_{2}$+2 jet production 
via vector boson fusion at NLO QCD accuracy.}
\vspace{0.2cm}
\label{tab:spin2}
\end{center}
\end{table}

\subsection{Final state summed over generations}

Besides decays into a single generation, results can now also be
calculated summed over different generations. This applies to both the
final-state leptons and, for the semileptonic decay modes, also the
final-state quarks appearing as decay product of vector bosons. The
cross section will then be summed over all possibilities, and Les
Houches event files will contain events with all included generations
and correct relative weights. This new option is controlled in
\texttt{vbfnlo.dat} by setting the final-state lepton and quark switches
\texttt{LEPTONS} and \texttt{DECAY\_QUARKS}, respectively. Setting the
former variable to \texttt{99} includes leptons of all three
generations, while \texttt{98} uses only those of the first and second
generation. The latter variable can be set to \texttt{93} to allow only
first and second generation decay products, or \texttt{94} to include
(massless) bottom quarks as well.

\subsection{Event output}

Some options have been added to make generating Les Houches event files
easier. It is now possible to specify a desired number of unweighted
events by setting \texttt{DESIRED\_EVENT\_COUNT} in \texttt{vbfnlo.dat}
to the corresponding value. If this number is not yet reached after
having completed the requested number of integration iterations, event
generation will continue until this is completed. 
For some processes or parameter settings it might be difficult to get
the desired number of unweighted events, because single events with
large weights spoil the unweighting efficiency. For this case
\texttt{PARTIAL\_UNWEIGHTING = true} will give the desired number of
events with unit weight and additionally a few events with weight $>1$
which occured during the event generation.

\section{Other changes}

Since the previous release, \textsc{Version 2.6.0}, some changes have been made
that alter previous results (events, cross sections and distributions).

\subsection{Running scales in NLO calculations}

In release 2.7.0 a problem with certain dynamical renormalization and factorization
scales has been fixed, which lead to wrong results at NLO QCD in the subtraction part of the
real emission calculation. In particular, the scale choices ``$\min(p_T(j_i))$'' (ID=2)
and ``minimal transverse energy of the bosons'' (ID=7) did not give correct results
in previous releases.

\subsection{Jet cuts in VBF/gluon fusion Higgs boson production with $H\to b\bar{b}$}

Starting with {\sc Vbfnlo 2.7.0} the jet cuts will be applied to the
$b$-quarks from Higgs boson decay.

\subsection{NLO calculation of $W^+W^-Z$ production}

A bug in the calculation of the virtual contributions for $WWZ$
production has been fixed while comparing with the results of
Ref.~\cite{Nhung:2013jta}.  The new results are roughly one per cent
smaller, and agree between both codes for squared amplitudes at the
level of the machine precision and for integrated cross sections at the
per mill level.

\subsection{LO and NLO calculation of $W^+jj$ / $Zjj$ production in VBF}

The particle--anti-particle--assignment in $W^+jj$, $Zjj$ has been fixed.
In case of $W^+jj$ production this leads to an increase in cross section of roughly one per cent with basic cuts.

\subsection{Event output}

Several bugs have been fixed concerning the event output to Les Houches or HepMC files:
\begin{itemize}
 \item The color information in the event output for $W^-W^- jj$ in VBF has been fixed.
 \item The momenta assignment in the event output for $W^-W^+W^-$ has been corrected.
\end{itemize}

\subsection{Previous changes -- version 2.6.3}

The release \textsc{Version 2.6.3} includes some changes that alter previous results:

\subsubsection{Event output}

Several bugs have been fixed concerning the event output to Les Houches or HepMC files:
\begin{itemize}
 \item Fixed bugs in output of particle IDs for processes 191 and 43xx: In previous versions
       some particles had the particle ID 0.
 \item Parton and beam particle IDs have been fixed for process 260.
 \item The tau mass can now be included in the event output of all processes. Furthermore,
       several bugs have been fixed in the existing implementation of tau mass inclusion.
 \item Several bugs have been fixed in the helicity output.
\end{itemize}

\subsubsection{Calculation of the $H\to gg$ partial width}

Higher-order corrections to the $H\to gg$ partial width have been included which lead to slightly smaller
branching ratios for all other decay channels. Therefore cross sections of processes involving Higgs bosons
can be up to a few per cent smaller.

\subsubsection{Electroweak corrections in the VBF $Hjj$ processes}

Some bugs have been fixed in the calculation of electroweak corrections for the processes 10x.

\subsection{Previous changes -- version 2.6.2}

The release \textsc{Version 2.6.2} included some changes that altered previous results:

\subsubsection{Distributions and cross section for $H\to WW$ in VBF and gluon fusion}

Due to a bug which has been fixed in the lepton assignment for $H\to WW/ZZ \to 4\ell$ distributions
(and cross sections after the $m_{\ell\ell}$ cut) were off for the 1x5 and 4105 processes
in the previous versions. This bug has been fixed in v.2.6.2.

\subsubsection{Les Houches event output for processes with more than one phase space}

The fraction of events coming from the different phase spaces was not sampled correctly 
in previous versions. Processes affected are $W\gamma$, $Z\gamma$, $W\gamma j$,
$Z\gamma j$, $ZZZ$, $WW\gamma$, $WZ\gamma$, $W\gamma\gamma$, $Z\gamma\gamma$,
$\gamma\gamma\gamma$ and $W\gamma\gamma j$.

\subsubsection{$ZZZ$ production}

The QCD real emission part of the NLO computation gave no reliable result in previous versions
due to a bug in the dipole subtraction. This bug has been fixed in v.2.6.2.

\subsubsection{Form factor in $\gamma\,jj$ production with anomalous couplings}

 For the $\gamma\,jj$ production in VBF (procID 150) a different form factor is used.

\subsection{Previous changes -- version 2.6.1}

The release \textsc{Version 2.6.1} included some changes that altered previous results:

\subsubsection{Anomalous Higgs couplings}

A bug was found and fixed in the implementation of the {\tt TREEFACZ} and {\tt TREEFACW}, 
the factors which multiply the SM $HZZ$ and $HWW$ couplings.  Note that this bug was only 
present in \textsc{Version 2.6.0}, not in earlier versions.  Additionally, a small bug was 
found and fixed in the coefficient of the input {\tt FB\_ODD} in the $a_{3}^{HZZ}$ coupling.

\subsubsection{Symmetry factor $ZZ \rightarrow \ell^{+} \ell^{-} \ell^{+} \ell^{-}$ }

In the processes $pp \rightarrow H \gamma jj \rightarrow ZZ \gamma jj \rightarrow \ell^{+} 
\ell^{-} \ell^{+} \ell^{-} \gamma jj$ (ID 2106) and gluon fusion $pp \rightarrow H jj \rightarrow 
ZZ jj \rightarrow \ell^{+} \ell^{-} \ell^{+} \ell^{-} jj $ (ID 4106) a symmetry factor was missing 
when identical final-state leptons were chosen.

%--------------------------------------------------------------------------------
%================= Section ======================================================
%--------------------------------------------------------------------------------

\section*{Acknowledgments}

We are grateful to Ken Arnold, Manuel B\"ahr, Guiseppe Bozzi, Martin Brieg,
Christoph Englert, Barbara J\"ager, Florian Geyer, Nicolas Greiner, Christoph
Hackstein, Vera Hankele, Nicolas Kaiser, Gunnar Kl\"amke, Michael Kubocz, Carlo
Oleari, Stefan Prestel, Heidi Rzehak, Michael Spannowsky and Malgorzata Worek
for their past contributions to the \textsc{Vbfnlo} code. 
We also gratefully acknowledge Simon Pl\"atzer's past contributions to the code
and his continued help on various technical aspects within \textsc{Vbfnlo}.
TF would like to thank the North American Foundation for The University of
Manchester and George Rigg for their financial support.

\providecommand{\href}[2]{#2}

\newpage
\section*{Appendix A: Process list}
\label{app:proc_list}
The following is a complete list of all processes available in \textsc{Vbfnlo},
including any Beyond the Standard Model (BSM) effects that are implemented. 
Firstly, the processes that are accessed via the {\tt vbfnlo} executable are given.
{
\footnotesize
\setlength\LTleft{0pt plus \textwidth minus \textwidth}
\setlength\LTright{0pt plus \textwidth minus \textwidth}
\begin{longtable}{clcccccccc}
\textsc{ProcId} & \textsc{Process} & \rot{semi-leptonic decay} & \rot{VBF process} & \rot{anom.\ gauge couplings} & \rot{anom.\ Higgs couplings} & \rot{Two-Higgs model} & \rot{Kaluza-Klein model} & \rot{Spin-2 model} & \rot{MSSM} \\
&\\
\hline
\endhead
&\\*
\bf 100 & $p \overset{\mbox{\tiny{(--)}}}{p} \to H \, jj$ &\bsmoptions{HMF}\\*
\bf 101 & $p \overset{\mbox{\tiny{(--)}}}{p} \to H \, jj\to \gamma\gamma \, jj$ &\bsmoptions{HMF}\\*
\bf 102 & $p \overset{\mbox{\tiny{(--)}}}{p} \to H \, jj\to \mu^+\mu^- \, jj$ &\bsmoptions{HMF}\\*
\bf 103 & $p \overset{\mbox{\tiny{(--)}}}{p} \to H \, jj\to \tau^+\tau^- \, jj$ &\bsmoptions{HMF}\\*
\bf 104 & $p \overset{\mbox{\tiny{(--)}}}{p} \to H \, jj\to b\bar{b} \, jj$ &\bsmoptions{HMF}\\*
\bf 105 & $p \overset{\mbox{\tiny{(--)}}}{p} \to H \, jj\to W^{+}W^{-} \, jj\to \ell_{1}^+\nu_{\ell_{1}} \ell_{2}^- \bar{\nu}_{\ell_{2}} \,jj$ &\bsmoptions{HMF}\\*
\bf 106 & $p \overset{\mbox{\tiny{(--)}}}{p} \to H \, jj\to ZZ \, jj\to \ell_{1}^+ \ell_{1}^- \ell_{2}^+ \ell_{2}^- \,jj$ &\bsmoptions{HMF}\\*
\bf 107 & $p \overset{\mbox{\tiny{(--)}}}{p} \to H \, jj\to ZZ \, jj\to \ell_{1}^+ \ell_{1}^- \nu_{\ell_{2}}  \bar{\nu}_{\ell_{2}} \,jj$ &\bsmoptions{HMF}\\*
%&\\
%\hline
%&\\
\bf 108 & $p \overset{\mbox{\tiny{(--)}}}{p} \to H \, jj\to W^{+}W^{-} \, jj\to q\bar{q} \, \ell^- \bar{\nu}_{\ell} \,jj$ &\bsmoptions{LHMF}\\*
\bf 109 & $p \overset{\mbox{\tiny{(--)}}}{p} \to H \, jj\to W^{+}W^{-} \, jj\to \ell^+\nu_{\ell} \, q\bar{q}  \,jj$ &\bsmoptions{LHMF}\\*
\bf 1010 & $p \overset{\mbox{\tiny{(--)}}}{p} \to H \, jj\to ZZ \, jj\to q\bar{q} \, \ell^+ \ell^- \,jj$ &\bsmoptions{LHMF}\\*
&\\*
\hline
&\\*
\bf 110 & $p \overset{\mbox{\tiny{(--)}}}{p} \to H \, jjj$ &\bsmoptions{F}\\*
\bf 111 & $p \overset{\mbox{\tiny{(--)}}}{p} \to H \, jjj\to \gamma\gamma \, jjj$ &\bsmoptions{F}\\*
\bf 112 & $p \overset{\mbox{\tiny{(--)}}}{p} \to H \, jjj\to \mu^+\mu^- \, jjj$ &\bsmoptions{F}\\*
\bf 113 & $p \overset{\mbox{\tiny{(--)}}}{p} \to H \, jjj\to \tau^+\tau^- \, jjj$ &\bsmoptions{F}\\*
\bf 114 & $p \overset{\mbox{\tiny{(--)}}}{p} \to H \, jjj\to b\bar{b} \, jjj$ &\bsmoptions{F}\\*
\bf 115 & $p \overset{\mbox{\tiny{(--)}}}{p} \to H \, jjj\to W^+W^- \, jjj\to \ell_{1}^{+}\nu_{\ell_{1}} \ell_{2}^- \bar{\nu}_{\ell_{2}} \,jjj$ &\bsmoptions{F}\\*
\bf 116 & $p \overset{\mbox{\tiny{(--)}}}{p} \to H \, jjj\to ZZ \, jjj\to \ell_{1}^+ \ell_{1}^- \ell_{2}^+ \ell_{2}^- \,jjj$ &\bsmoptions{F}\\*
\bf 117 & $p \overset{\mbox{\tiny{(--)}}}{p} \to H \, jjj\to ZZ \, jjj\to \ell_{1}^+ \ell_{1}^- \nu_{\ell_{2}} \bar{\nu}_{\ell_{2}} \,jjj$ &\bsmoptions{F}\\*
&\\*
\hline
&\\*
\bf 120 & $p \overset{\mbox{\tiny{(--)}}}{p} \to Z \, jj \to \ell^{+} \ell^{-} \, jj$ &\bsmoptions{VF}\\*
\bf 121 & $p \overset{\mbox{\tiny{(--)}}}{p} \to Z  \, jj\to \nu_\ell \bar{\nu}_\ell \, jj$ &\bsmoptions{VF}\\*
\bf 130 & $p \overset{\mbox{\tiny{(--)}}}{p} \to W^{+} \,  jj\to \ell^{+} \nu_\ell \, jj$ &\bsmoptions{VF}\\*
\bf 140 & $p \overset{\mbox{\tiny{(--)}}}{p} \to W^{-} \, jj\to \ell^{-} \bar{\nu}_\ell  \, jj$ &\bsmoptions{VF}\\*
\bf 150 & $p \overset{\mbox{\tiny{(--)}}}{p} \to \gamma \, jj$ &\bsmoptions{VF}\\*
&\\*
\hline
&\\*
\bf 191 & $p \overset{\mbox{\tiny{(--)}}}{p} \to S_{2}  \, jj\to \gamma \gamma \, jj$ &\bsmoptions{SF}\\*
\bf 195 & $p \overset{\mbox{\tiny{(--)}}}{p} \to S_{2}  \, jj\to W^{+}W^{-} \, jj\to \ell_{1}^+\nu_{\ell_{1}} \ell_{2}^- \bar{\nu}_{\ell_{2}} \,jj$ &\bsmoptions{SF}\\*
\bf 196 & $p \overset{\mbox{\tiny{(--)}}}{p} \to S_{2}  \, jj\to ZZ \, jj\to \ell_{1}^+ \ell_{1}^- \ell_{2}^+ \ell_{2}^- \,jj$ &\bsmoptions{SF}\\*
\bf 197 & $p \overset{\mbox{\tiny{(--)}}}{p} \to S_{2}  \, jj\to  ZZ \, jj\to \ell_{1}^+ \ell_{1}^- \nu_{\ell_{2}} \bar{\nu}_{\ell_{2}} \,jj$ &\bsmoptions{SF}\\*
&\\*
\hline
&\\*
\bf 160 & $p \overset{\mbox{\tiny{(--)}}}{p} \to HH \,  jj$ &\bsmoptions{F}\\*
&\\*
\hline
&\\*
\bf 2100 & $p \overset{\mbox{\tiny{(--)}}}{p} \to H \gamma \, jj$ &\bsmoptions{F}\\*
\bf 2101 & $p \overset{\mbox{\tiny{(--)}}}{p} \to H \gamma \, jj\to \gamma\gamma \gamma \, jj$ &\bsmoptions{F}\\*
\bf 2102 & $p \overset{\mbox{\tiny{(--)}}}{p} \to H \gamma \, jj\to \mu^+\mu^- \gamma \, jj$ &\bsmoptions{F}\\*
\bf 2103 & $p \overset{\mbox{\tiny{(--)}}}{p} \to H \gamma \, jj\to \tau^+\tau^- \gamma \, jj$ &\bsmoptions{F}\\*
\bf 2104 & $p \overset{\mbox{\tiny{(--)}}}{p} \to H \gamma \, jj\to b\bar{b} \gamma \, jj$ &\bsmoptions{F}\\*
\bf 2105 & $p \overset{\mbox{\tiny{(--)}}}{p} \to H \gamma \, jj\to W^+W^- \gamma \, jj\to \ell_{1}^+\nu_{\ell_{1}} \ell_{2}^- \bar{\nu}_{\ell_{2}} \gamma \,jj$ &\bsmoptions{F}\\*
\bf 2106 & $p \overset{\mbox{\tiny{(--)}}}{p} \to H \gamma \, jj\to ZZ \gamma \, jj\to \ell_{1}^+ \ell_{1}^- \ell_{2}^+ \ell_{2}^- \gamma \,jj$ &\bsmoptions{F}\\*
\bf 2107 & $p \overset{\mbox{\tiny{(--)}}}{p} \to H \gamma \, jj\to ZZ \gamma \, jj\to \ell_{1}^+ \ell_{1}^- \nu_{\ell_{2}}  \bar{\nu}_{\ell_{2}} \gamma \,jj$ &\bsmoptions{F}\\*
&\\*
\hline
&\\*
\bf 200 & $p \overset{\mbox{\tiny{(--)}}}{p} \to W^{+}W^{-} \, jj \to \ell_{1}^{+} \nu_{\ell_{1}} \ell_{2}^{-} \bar{\nu}_{\ell_{2}} \, jj$ &\bsmoptions{VTKSF}\\*
\bf 201 & $p \overset{\mbox{\tiny{(--)}}}{p} \to W^{+}W^{-} \, jj \to q \bar{q} \, \ell^{-} \bar{\nu}_{\ell} \, jj$ &\bsmoptions{VTLF}\\*
\bf 202 & $p \overset{\mbox{\tiny{(--)}}}{p} \to W^{+}W^{-}  \, jj\to \ell^{+} \nu_\ell \, q \bar{q} \, jj$ &\bsmoptions{VTLF}\\*
\bf 210 & $p \overset{\mbox{\tiny{(--)}}}{p} \to ZZ  \, jj\to \ell_{1}^{+} \ell_{1}^{-} \ell_{2}^{+} \ell_{2}^{-} \, jj$ &\bsmoptions{VTKSF}\\*
\bf 211 & $p \overset{\mbox{\tiny{(--)}}}{p} \to ZZ  \, jj\to \ell_{1}^{+} \ell_{1}^{-} \nu_{\ell_{2}} \bar{\nu}_{\ell_{2}} \, jj$ &\bsmoptions{VTKSF}\\*
\bf 212 & $p \overset{\mbox{\tiny{(--)}}}{p} \to ZZ  \, jj\to q \bar{q} \, \ell^{+} \ell^{-} \, jj$ &\bsmoptions{VTLF}\\*
\bf 220 & $p \overset{\mbox{\tiny{(--)}}}{p} \to W^{+}Z \,  jj\to \ell_{1}^{+} \nu_{\ell_{1}} \ell_{2}^{+} \ell_{2}^{-} \, jj$ &\bsmoptions{VTKSF}\\*
\bf 221 & $p \overset{\mbox{\tiny{(--)}}}{p} \to W^{+}Z \,  jj\to q \bar{q} \, \ell^{+} \ell^{-} \, jj$ &\bsmoptions{VTLF}\\*
\bf 222 & $p \overset{\mbox{\tiny{(--)}}}{p} \to W^{+}Z \,  jj\to \ell^{+} \nu_{\ell} \, q \bar{q} \, jj$ &\bsmoptions{VTLF}\\*
\bf 230 & $p \overset{\mbox{\tiny{(--)}}}{p} \to W^{-}Z \, jj\to \ell_{1}^{-} \bar{\nu}_{\ell _{1}} \ell_{2}^{+} \ell_{2}^{-} \, jj$ &\bsmoptions{VTKSF}\\*
\bf 231 & $p \overset{\mbox{\tiny{(--)}}}{p} \to W^{-}Z \, jj\to q \bar{q} \, \ell^{+} \ell^{-} \, jj$ &\bsmoptions{VTLF}\\*
\bf 232 & $p \overset{\mbox{\tiny{(--)}}}{p} \to W^{-}Z \, jj\to \ell^{-} \bar{\nu}_{\ell} \, q \bar{q} \, jj$ &\bsmoptions{VTLF}\\*
\bf 250 & $p \overset{\mbox{\tiny{(--)}}}{p} \to W^{+}W^{+} \,  jj\to \ell_{1}^{+} \nu_{\ell_{1}} \ell_{2}^{+} \nu_{\ell_{2}} \, jj$ &\bsmoptions{VTF}\\*
\bf 251 & $p \overset{\mbox{\tiny{(--)}}}{p} \to W^{+}W^{+} \,  jj\to q \bar{q} \, \ell^{+} \nu_{\ell} \, jj$ &\bsmoptions{VTLF}\\*
\bf 260 & $p \overset{\mbox{\tiny{(--)}}}{p} \to W^{-}W^{-} \,  jj\to \ell_{1}^{-} \bar{\nu}_{\ell_{1}} \ell_{2}^{-} \bar{\nu}_{\ell_{2}} \, jj$ &\bsmoptions{VTF}\\*
\bf 261 & $p \overset{\mbox{\tiny{(--)}}}{p} \to W^{-}W^{-} \,  jj\to q \bar{q} \, \ell^{-} \bar{\nu}_{\ell} \, jj$ &\bsmoptions{VTLF}\\*
\bf 270 & $p \overset{\mbox{\tiny{(--)}}}{p} \to W^{+}\gamma \, jj\to \ell^{+} \nu_{\ell} \gamma \, jj$ &\bsmoptions{VF}\\*
\bf 280 & $p \overset{\mbox{\tiny{(--)}}}{p} \to W^{-}\gamma \, jj\to \ell^{-} \bar{\nu}_{\ell} \gamma \, jj$ &\bsmoptions{VF}\\*
&\\*
\hline
&\\*
\bf 3130 & $p \overset{\mbox{\tiny{(--)}}}{p} \to W^{+} \,  jj\to \ell^{+} \nu_\ell \, jj$ &\bsmoptions{}\\*
\bf 3140 & $p \overset{\mbox{\tiny{(--)}}}{p} \to W^{-} \, jj\to \ell^{-} \bar{\nu}_\ell  \, jj$ &\bsmoptions{}\\*
&\\*
\hline
&\\*
\bf 3220 & $p \overset{\mbox{\tiny{(--)}}}{p} \to W^{+}Z \,  jj\to \ell_{1}^{+} \nu_{\ell_{1}} \ell_{2}^{+} \ell_{2}^{-} \, jj$ &\bsmoptions{}\\*
\bf 3230 & $p \overset{\mbox{\tiny{(--)}}}{p} \to W^{-}Z \, jj\to \ell_{1}^{-} \bar{\nu}_{\ell _{1}} \ell_{2}^{+} \ell_{2}^{-} \, jj$ &\bsmoptions{}\\*
\bf 3250 & $p \overset{\mbox{\tiny{(--)}}}{p} \to W^{+}W^{+} \,  jj\to \ell_{1}^{+} \nu_{\ell_{1}} \ell_{2}^{+} \nu_{\ell_{2}} \, jj$ &\bsmoptions{}\\*
\bf 3260 & $p \overset{\mbox{\tiny{(--)}}}{p} \to W^{-}W^{-} \,  jj\to \ell_{1}^{-} \bar{\nu}_{\ell_{1}} \ell_{2}^{-} \bar{\nu}_{\ell_{2}} \, jj$ &\bsmoptions{}\\*
\bf 3270 & $p \overset{\mbox{\tiny{(--)}}}{p} \to W^{+}\gamma \,  jj\to \ell^{+} \nu_{\ell} \gamma \, jj$ &\bsmoptions{}\\*
\bf 3280 & $p \overset{\mbox{\tiny{(--)}}}{p} \to W^{-}\gamma \, jj\to \ell^{-} \bar{\nu}_{\ell} \gamma \, jj$ &\bsmoptions{}\\*
&\\*
\hline
&\\*
\bf 1330 & $p \overset{\mbox{\tiny{(--)}}}{p} \to W^+ \to \ell^+\nu_{\ell} $ &\bsmoptions{}\\
\bf 1340 & $p \overset{\mbox{\tiny{(--)}}}{p} \to W^- \to \ell^- \bar{\nu}_{\ell} $ &\bsmoptions{}\\
\bf 1630 & $p \overset{\mbox{\tiny{(--)}}}{p} \to W^+ \, j \to \ell^+\nu_{\ell} \, j $ &\bsmoptions{}\\
\bf 1640 & $p \overset{\mbox{\tiny{(--)}}}{p} \to W^- \, j\to \ell^- \bar{\nu}_{\ell} \, j $ &\bsmoptions{}\\
&\\*
\hline
&\\*
\bf 300 & $p \overset{\mbox{\tiny{(--)}}}{p} \to W^{+}W^{-} \to \ell_{1}^{+} \nu_{\ell_{1}} \ell_{2}^{-}\bar{\nu}_{\ell_{2}} $ &\bsmoptions{HV}\\*
\bf 301 & $p \overset{\mbox{\tiny{(--)}}}{p} \to W^{+}W^{-} \to q \bar{q} \, \ell^{-}\bar{\nu}_{\ell} $ &\bsmoptions{HVL}\\*
\bf 302 & $p \overset{\mbox{\tiny{(--)}}}{p} \to W^{+}W^{-} \to \ell^{+} \nu_{\ell} \, q \bar{q} $ &\bsmoptions{HVL}\\*
\bf 310 & $p \overset{\mbox{\tiny{(--)}}}{p} \to W^{+}Z \to  \ell_{1}^{+} \nu_{\ell_1}  \ell_{2}^{+} \ell_{2}^{-} $ &\bsmoptions{V}\\*
\bf 312 & $p \overset{\mbox{\tiny{(--)}}}{p} \to W^{+}Z \to  q \bar{q} \, \ell^{+} \ell^{-} $ &\bsmoptions{VL}\\*
\bf 313 & $p \overset{\mbox{\tiny{(--)}}}{p} \to W^{+}Z \to  \ell^{+} \nu_{\ell} \, q \bar{q} $ &\bsmoptions{VL}\\*
\bf 320 & $p \overset{\mbox{\tiny{(--)}}}{p} \to W^{-}Z \to \ell_{1}^{-} \bar{\nu}_{\ell_{1}}  \ell_{2}^{+} \ell_{2}^{-} $ &\bsmoptions{V}\\*
\bf 322 & $p \overset{\mbox{\tiny{(--)}}}{p} \to W^{-}Z \to q \bar{q} \, \ell^{+} \ell^{-} $ &\bsmoptions{VL}\\*
\bf 323 & $p \overset{\mbox{\tiny{(--)}}}{p} \to W^{-}Z \to \ell^{-} \bar{\nu}_{\ell} \, q \bar{q} $ &\bsmoptions{VL}\\*
\bf 330 & $p \overset{\mbox{\tiny{(--)}}}{p} \to ZZ \to \ell_{1}^{-} \ell_{1}^{+}  \ell_{2}^{-} \ell_{2}^{+} $ &\bsmoptions{H}\\*
\bf 331 & $p \overset{\mbox{\tiny{(--)}}}{p} \to ZZ \to q \bar{q} \, \ell^{-} \ell^{+} $ &\bsmoptions{HL}\\*
\bf 340 & $p \overset{\mbox{\tiny{(--)}}}{p} \to W^{+}\gamma \to \ell_{1}^{+} \nu_{\ell_1} \gamma $ &\bsmoptions{V}\\*
\bf 350 & $p \overset{\mbox{\tiny{(--)}}}{p} \to W^{-}\gamma \to \ell_{1}^{-} \bar{\nu}_{\ell_{1}} \gamma $ &\bsmoptions{V}\\*
\bf 360 & $p \overset{\mbox{\tiny{(--)}}}{p} \to Z\gamma \to \ell_{1}^{-} \ell_{1}^{+}  \gamma $ &\bsmoptions{H}\\*
\bf 370 & $p \overset{\mbox{\tiny{(--)}}}{p} \to \gamma\gamma $ &\bsmoptions{H}\\*
&\\*
\hline
&\\*
\bf 1300 & $p \overset{\mbox{\tiny{(--)}}}{p} \to W^+ H \to \ell^+\nu_{\ell} H $ &\bsmoptions{V}\\*
\bf 1301 & $p \overset{\mbox{\tiny{(--)}}}{p} \to W^+ H \to \ell^+\nu_{\ell} \gamma\gamma $ &\bsmoptions{V}\\*
\bf 1302 & $p \overset{\mbox{\tiny{(--)}}}{p} \to W^+ H \to \ell^+\nu_{\ell} \mu^+\mu^- $ &\bsmoptions{V}\\*
\bf 1303 & $p \overset{\mbox{\tiny{(--)}}}{p} \to W^+ H \to \ell^+\nu_{\ell} \tau^+\tau^- $ &\bsmoptions{V}\\*
\bf 1304 & $p \overset{\mbox{\tiny{(--)}}}{p} \to W^+ H \to \ell^+\nu_{\ell} b\bar{b} $ &\bsmoptions{V}\\*
\bf 1305 & $p \overset{\mbox{\tiny{(--)}}}{p} \to W^+ H \to W^+ W^{+}W^{-} \to \ell_{1}^+\nu_{\ell_{1}} \ell_{2}^+\nu_{\ell_{2}} \ell_{3}^- \bar{\nu}_{\ell_{3}}$ &\bsmoptions{V}\\*
\bf 1306 & $p \overset{\mbox{\tiny{(--)}}}{p} \to W^+ H \to W^+ ZZ \to \ell_{1}^+\nu_{\ell_{1}} \ell_{2}^+ \ell_{2}^- \ell_{3}^+ \ell_{3}^-$ &\bsmoptions{V}\\*
\bf 1307 & $p \overset{\mbox{\tiny{(--)}}}{p} \to W^+ H \to W^+ ZZ \to \ell_{1}^+\nu_{\ell_{1}} \ell_{2}^+ \ell_{2}^- \nu_{\ell_{3}}  \bar{\nu}_{\ell_{3}}$ &\bsmoptions{V}\\*
&\\*
\hline
&\\*
\bf 1310 & $p \overset{\mbox{\tiny{(--)}}}{p} \to W^- H \to \ell^- \bar{\nu}_{\ell} H $ &\bsmoptions{V}\\*
\bf 1311 & $p \overset{\mbox{\tiny{(--)}}}{p} \to W^- H \to \ell^- \bar{\nu}_{\ell} \gamma\gamma $ &\bsmoptions{V}\\*
\bf 1312 & $p \overset{\mbox{\tiny{(--)}}}{p} \to W^- H \to \ell^- \bar{\nu}_{\ell} \mu^+\mu^- $ &\bsmoptions{V}\\*
\bf 1313 & $p \overset{\mbox{\tiny{(--)}}}{p} \to W^- H \to \ell^- \bar{\nu}_{\ell} \tau^+\tau^- $ &\bsmoptions{V}\\*
\bf 1314 & $p \overset{\mbox{\tiny{(--)}}}{p} \to W^- H \to \ell^- \bar{\nu}_{\ell} b\bar{b} $ &\bsmoptions{V}\\*
\bf 1315 & $p \overset{\mbox{\tiny{(--)}}}{p} \to W^- H \to W^- W^{+}W^{-} \to \ell_{1}^-\bar{\nu}_{\ell_{1}} \ell_{2}^+\nu_{\ell_{2}} \ell_{3}^- \bar{\nu}_{\ell_{3}}$ &\bsmoptions{V}\\*
\bf 1316 & $p \overset{\mbox{\tiny{(--)}}}{p} \to W^- H \to W^- ZZ \to \ell_{1}^-\bar{\nu}_{\ell_{1}} \ell_{2}^+ \ell_{2}^- \ell_{3}^+ \ell_{3}^-$ &\bsmoptions{V}\\*
\bf 1317 & $p \overset{\mbox{\tiny{(--)}}}{p} \to W^- H \to W^- ZZ \to \ell_{1}^-\bar{\nu}_{\ell_{1}} \ell_{2}^+ \ell_{2}^- \nu_{\ell_{3}}  \bar{\nu}_{\ell_{3}}$ &\bsmoptions{V}\\*
&\\*
\hline
&\\*
\bf 610 & $p \overset{\mbox{\tiny{(--)}}}{p}  \to W^{-} \gamma j \to \ell^{-} \bar \nu_{\ell} \gamma j $ &\bsmoptions{V}\\*
\bf 620 & $p \overset{\mbox{\tiny{(--)}}}{p}  \to W^{+} \gamma j  \to \ell^{+} \nu_{\ell} \gamma j $ &\bsmoptions{V}\\*
\bf 630 & $p \overset{\mbox{\tiny{(--)}}}{p}  \to W^{-} Z j \to \ell_{1}^{-} \bar \nu_{\ell_1} \ell_{2}^{-} \ell_{2}^{+} j$ &\bsmoptions{V}\\*
\bf 640 & $p \overset{\mbox{\tiny{(--)}}}{p}  \to W^{+} Z j \to  \ell_{1}^{+}\nu_{\ell_1} \ell_{2}^{-} \ell_{2}^{+}j $ &\bsmoptions{V}\\*
&\\*
\hline
&\\*
\bf 1600 & $p \overset{\mbox{\tiny{(--)}}}{p} \to W^+ H \, j \to \ell^+\nu_{\ell} H \, j $ &\bsmoptions{V}\\*
\bf 1601 & $p \overset{\mbox{\tiny{(--)}}}{p} \to W^+ H \, j \to \ell^+\nu_{\ell} \gamma\gamma \, j $ &\bsmoptions{V}\\*
\bf 1602 & $p \overset{\mbox{\tiny{(--)}}}{p} \to W^+ H \, j \to \ell^+\nu_{\ell} \mu^+\mu^- \, j $ &\bsmoptions{V}\\*
\bf 1603 & $p \overset{\mbox{\tiny{(--)}}}{p} \to W^+ H \, j \to \ell^+\nu_{\ell} \tau^+\tau^- \, j $ &\bsmoptions{V}\\*
\bf 1604 & $p \overset{\mbox{\tiny{(--)}}}{p} \to W^+ H \, j \to \ell^+\nu_{\ell} b\bar{b} \, j $ &\bsmoptions{V}\\*
\bf 1605 & $p \overset{\mbox{\tiny{(--)}}}{p} \to W^+ H \, j \to W^+ W^{+}W^{-} \, j \to \ell_{1}^+\nu_{\ell_{1}} \ell_{2}^+\nu_{\ell_{2}} \ell_{3}^- \bar{\nu}_{\ell_{3}}\, j $ &\bsmoptions{V}\\*
\bf 1606 & $p \overset{\mbox{\tiny{(--)}}}{p} \to W^+ H \, j \to W^+ ZZ \, j \to \ell_{1}^+\nu_{\ell_{1}} \ell_{2}^+ \ell_{2}^- \ell_{3}^+ \ell_{3}^-\, j $ &\bsmoptions{V}\\*
\bf 1607 & $p \overset{\mbox{\tiny{(--)}}}{p} \to W^+ H \, j \to W^+ ZZ \, j \to \ell_{1}^+\nu_{\ell_{1}} \ell_{2}^+ \ell_{2}^- \nu_{\ell_{3}}  \bar{\nu}_{\ell_{3}}\, j $ &\bsmoptions{V}\\*
&\\*
\hline
&\\*
\bf 1610 & $p \overset{\mbox{\tiny{(--)}}}{p} \to W^- H \, j \to \ell^- \bar{\nu}_{\ell} H \, j $ &\bsmoptions{V}\\*
\bf 1611 & $p \overset{\mbox{\tiny{(--)}}}{p} \to W^- H \, j \to \ell^- \bar{\nu}_{\ell} \gamma\gamma \, j $ &\bsmoptions{V}\\*
\bf 1612 & $p \overset{\mbox{\tiny{(--)}}}{p} \to W^- H \, j \to \ell^- \bar{\nu}_{\ell} \mu^+\mu^- \, j $ &\bsmoptions{V}\\*
\bf 1613 & $p \overset{\mbox{\tiny{(--)}}}{p} \to W^- H \, j \to \ell^- \bar{\nu}_{\ell} \tau^+\tau^- \, j $ &\bsmoptions{V}\\*
\bf 1614 & $p \overset{\mbox{\tiny{(--)}}}{p} \to W^- H \, j \to \ell^- \bar{\nu}_{\ell} b\bar{b} \, j $ &\bsmoptions{V}\\*
\bf 1615 & $p \overset{\mbox{\tiny{(--)}}}{p} \to W^- H \, j \to W^- W^{+}W^{-} \, j \to \ell_{1}^-\bar{\nu}_{\ell_{1}} \ell_{2}^+\nu_{\ell_{2}} \ell_{3}^- \bar{\nu}_{\ell_{3}}\, j $ &\bsmoptions{V}\\*
\bf 1616 & $p \overset{\mbox{\tiny{(--)}}}{p} \to W^- H \, j \to W^- ZZ \, j \to \ell_{1}^-\bar{\nu}_{\ell_{1}} \ell_{2}^+ \ell_{2}^- \ell_{3}^+ \ell_{3}^-\, j $ &\bsmoptions{V}\\*
\bf 1617 & $p \overset{\mbox{\tiny{(--)}}}{p} \to W^- H \, j \to W^- ZZ \, j \to \ell_{1}^-\bar{\nu}_{\ell_{1}} \ell_{2}^+ \ell_{2}^- \nu_{\ell_{3}}  \bar{\nu}_{\ell_{3}}\, j $ &\bsmoptions{V}\\*
&\\*
\hline
&\\*
\bf 400 & $p \overset{\mbox{\tiny{(--)}}}{p} \to W^{+}W^{-}Z \to \ell_{1}^{+}\nu_{\ell_{1}} \ell_{2}^{-} \bar{\nu}_{\ell_{2}} \ell_{3}^{+} \ell_{3}^{-} $ &\bsmoptions{VK}\\*
\bf 401 & $p \overset{\mbox{\tiny{(--)}}}{p} \to W^{+}W^{-}Z \to q \bar{q} \, \ell_{1}^{-} \bar{\nu}_{\ell_{1}} \ell_{2}^{+} \ell_{2}^{-} $ &\bsmoptions{LV}\\*
\bf 402 & $p \overset{\mbox{\tiny{(--)}}}{p} \to W^{+}W^{-}Z \to \ell_{1}^{+}\nu_{\ell_{1}} \, q \bar{q} \, \ell_{2}^{+} \ell_{2}^{-} $ &\bsmoptions{LV}\\*
\bf 403 & $p \overset{\mbox{\tiny{(--)}}}{p} \to W^{+}W^{-}Z \to \ell_{1}^{+}\nu_{\ell_{1}} \ell_{2}^{-} \bar{\nu}_{\ell_{2}} \, q \bar{q} $ &\bsmoptions{LV}\\*
\bf 410 & $p \overset{\mbox{\tiny{(--)}}}{p} \to ZZW^{+} \to  \ell_{1}^{+} \ell_{1}^{-}  \ell_{2}^{+} \ell_{2}^{-} \ell_{3}^{+} \nu_{\ell_{3}} $ &\bsmoptions{VK}\\*
\bf 411 & $p \overset{\mbox{\tiny{(--)}}}{p} \to ZZW^{+} \to  \ell_{1}^{+} \ell_{1}^{-}  \ell_{2}^{+} \ell_{2}^{-} \, q \bar{q} $ &\bsmoptions{LV}\\*
\bf 412 & $p \overset{\mbox{\tiny{(--)}}}{p} \to ZZW^{+} \to  q \bar{q} \, \ell_{1}^{+} \ell_{1}^{-} \ell_{2}^{+} \nu_{\ell_{2}} $ &\bsmoptions{LV}\\*
\bf 420 & $p \overset{\mbox{\tiny{(--)}}}{p} \to ZZW^{-} \to \ell_{1}^{+} \ell_{1}^{-}  \ell_{2}^{+} \ell_{2}^{-} \ell_{3}^{-}  \bar{\nu}_{\ell_{3}}$ &\bsmoptions{VK}\\*
\bf 421 & $p \overset{\mbox{\tiny{(--)}}}{p} \to ZZW^{-} \to \ell_{1}^{+} \ell_{1}^{-}  \ell_{2}^{+} \ell_{2}^{-} \, q \bar{q} $ &\bsmoptions{LV}\\*
\bf 422 & $p \overset{\mbox{\tiny{(--)}}}{p} \to ZZW^{-} \to q \bar{q} \, \ell_{1}^{+} \ell_{1}^{-} \ell_{2}^{-}  \bar{\nu}_{\ell_{2}}$ &\bsmoptions{LV}\\*
\bf 430 & $p \overset{\mbox{\tiny{(--)}}}{p} \to W^{+}W^{-}W^{+} \to \ell_{1}^{+}\nu_{\ell_1} \ell_{2}^{-} \bar{\nu}_{\ell_2} \ell_{3}^{+}\nu_{\ell_{3}}$ &\bsmoptions{VK}\\*
\bf 431 & $p \overset{\mbox{\tiny{(--)}}}{p} \to W^{+}W^{-}W^{+} \to  q \bar{q} \, \ell_{1}^{-} \bar{\nu}_{\ell_1} \ell_{2}^{+}\nu_{\ell_{2}}$ &\bsmoptions{LV}\\*
\bf 432 & $p \overset{\mbox{\tiny{(--)}}}{p} \to W^{+}W^{-}W^{+} \to \ell_{1}^{+}\nu_{\ell_1} \, q \bar{q} \, \ell_{2}^{+}\nu_{\ell_{2}}$ &\bsmoptions{LV}\\*
\bf 440 & $p \overset{\mbox{\tiny{(--)}}}{p} \to W^{-}W^{+}W^{-} \to \ell_{1}^{-} \bar{\nu}_{\ell_1}\ell_{2}^{+}\nu_{\ell_2} \ell_{3}^{-} \bar{\nu}_{\ell_{3}} $ &\bsmoptions{VK}\\*
\bf 441 & $p \overset{\mbox{\tiny{(--)}}}{p} \to W^{-}W^{+}W^{-} \to \ell_{1}^{-} \bar{\nu}_{\ell_1} \, q \bar{q} \, \ell_{2}^{-} \bar{\nu}_{\ell_{2}} $ &\bsmoptions{LV}\\*
\bf 442 & $p \overset{\mbox{\tiny{(--)}}}{p} \to W^{-}W^{+}W^{-} \to  q \bar{q} \, \ell_{1}^{+}\nu_{\ell_1} \ell_{2}^{-} \bar{\nu}_{\ell_{2}} $ &\bsmoptions{LV}\\*
\bf 450 & $p \overset{\mbox{\tiny{(--)}}}{p} \to ZZZ \to \ell_{1}^{-} \ell_{1}^{+} \ell_{2}^{-} \ell_{2}^{+} \ell_{3}^{-} \ell_{3}^{+} $ &\bsmoptions{V}\\*
\bf 451 & $p \overset{\mbox{\tiny{(--)}}}{p} \to ZZZ \to  q \bar{q} \, \ell_{1}^{-} \ell_{1}^{+} \ell_{2}^{-} \ell_{2}^{+}  $ &\bsmoptions{LV}\\*
&\\*
\hline
&\\*
\bf 460 & $p \overset{\mbox{\tiny{(--)}}}{p} \to W^{-}W^{+} \gamma \to \ell_{1}^{-} \bar{\nu}_{\ell_1} \ell_{2}^{+}\nu_{\ell_2} \gamma$ &\bsmoptions{V}\\*
\bf 461 & $p \overset{\mbox{\tiny{(--)}}}{p} \to W^{+}W^{-} \gamma \to  q \bar{q} \, \ell^{-}\bar{\nu}_{\ell} \gamma$ &\bsmoptions{LV}\\*
\bf 462 & $p \overset{\mbox{\tiny{(--)}}}{p} \to W^{+}W^{-} \gamma \to \ell^{+} \nu_{\ell} \, q \bar{q} \, \gamma$ &\bsmoptions{LV}\\*
\bf 470 & $p \overset{\mbox{\tiny{(--)}}}{p} \to Z Z \gamma \to \ell_{1}^{-} \ell_{1}^{+} \ell_{2}^{-} \ell_{2}^{+} \gamma$ &\bsmoptions{V}\\*
\bf 471 & $p \overset{\mbox{\tiny{(--)}}}{p} \to Z Z \gamma \to \ell^{-} \ell^{+} \, q \bar{q} \, \gamma$ &\bsmoptions{LV}\\*
\bf 480 & $p \overset{\mbox{\tiny{(--)}}}{p} \to W^{+} Z \gamma \to \ell_{1}^{+}\nu_{\ell_1} \ell_{2}^{-} \ell_{2}^{+} \gamma$ &\bsmoptions{V}\\*
\bf 481 & $p \overset{\mbox{\tiny{(--)}}}{p} \to W^{+} Z \gamma \to  q \bar{q} \, \ell^{-} \ell^{+} \gamma$ &\bsmoptions{LV}\\*
\bf 482 & $p \overset{\mbox{\tiny{(--)}}}{p} \to W^{+} Z \gamma \to \ell^{+}\nu_{\ell} \, q \bar{q} \, \gamma$ &\bsmoptions{LV}\\*
\bf 490 & $p \overset{\mbox{\tiny{(--)}}}{p} \to W^{-} Z \gamma \to \ell_{1}^{-} \bar{\nu}_{\ell_1} \ell_{2}^{-} \ell_{2}^{+} \gamma$ &\bsmoptions{V}\\*
\bf 491 & $p \overset{\mbox{\tiny{(--)}}}{p} \to W^{-} Z \gamma \to  q \bar{q} \, \ell^{-} \ell^{+} \gamma$ &\bsmoptions{LV}\\*
\bf 492 & $p \overset{\mbox{\tiny{(--)}}}{p} \to W^{-} Z \gamma \to \ell^{-} \bar{\nu}_{\ell} \, q \bar{q} \, \gamma$ &\bsmoptions{LV}\\*
\bf 500 & $p \overset{\mbox{\tiny{(--)}}}{p} \to W^{+} \gamma \gamma \to {\ell}^{+}\nu_{\ell} \gamma \gamma$ &\bsmoptions{V}\\*
\bf 510 & $p \overset{\mbox{\tiny{(--)}}}{p} \to W^{-} \gamma \gamma \to {\ell}^{-} \bar{\nu}_{\ell} \gamma \gamma$ &\bsmoptions{V}\\*
\bf 520 & $p \overset{\mbox{\tiny{(--)}}}{p} \to Z \gamma \gamma \to {\ell}^{-} {\ell}^{+} \gamma \gamma$ &\bsmoptions{V}\\*
\bf 521 & $p \overset{\mbox{\tiny{(--)}}}{p} \to Z \gamma \gamma \to \nu_{\ell} \bar{\nu}_{\ell} \gamma \gamma$ &\bsmoptions{V}\\*
\bf 530 & $p \overset{\mbox{\tiny{(--)}}}{p} \to \gamma \gamma \gamma $ &\bsmoptions{V}\\*
&\\*
\hline
&\\*
\bf 800 & $p \overset{\mbox{\tiny{(--)}}}{p}  \to W^{+} \gamma \gamma j  \to \ell^{+} \nu_{\ell} \gamma \gamma j $ &\bsmoptions{V}\\*
\bf 810 & $p \overset{\mbox{\tiny{(--)}}}{p}  \to W^{-} \gamma \gamma j \to \ell^{-} \bar \nu_{\ell} \gamma \gamma j $ &\bsmoptions{V}\\*
&\\*
\hline
\end{longtable}
}

\clearpage
The gluon-fusion processes accessed via the executable {\tt ggflo} are given below.

{
\footnotesize
\setlength\LTleft{0pt plus \textwidth minus \textwidth}
\setlength\LTright{0pt plus \textwidth minus \textwidth}
\begin{longtable}{clcccccccc}
\textsc{ProcId} & \textsc{Process} & \rot{gluon-fusion process} & \rot{semi-leptonic decay} & \rot{anom.\ Higgs couplings} & \rot{general 2HDM} & \rot{MSSM} \\
&\\
\hline
\endhead
&\\*
\bf 4100 & $p \overset{\mbox{\tiny{(--)}}}{p} \to H \, jj $ &\bsmgfoptions{GTM}\\*
\bf 4101 & $p \overset{\mbox{\tiny{(--)}}}{p} \to H \, jj\to \gamma\gamma \, jj$ &\bsmgfoptions{GM}\\*
\bf 4102 & $p \overset{\mbox{\tiny{(--)}}}{p} \to H \, jj\to \mu^+\mu^- \, jj$ &\bsmgfoptions{GM}\\*
\bf 4103 & $p \overset{\mbox{\tiny{(--)}}}{p} \to H \, jj\to \tau^+\tau^- \, jj$ &\bsmgfoptions{GM}\\*
\bf 4104 & $p \overset{\mbox{\tiny{(--)}}}{p} \to H \, jj\to b\bar{b} \, jj$ &\bsmgfoptions{GM}\\*
\bf 4105 & $p \overset{\mbox{\tiny{(--)}}}{p} \to H \, jj\to W^{+}W^{-} \, jj\to \ell_{1}^+\nu_{\ell_{1}} \ell_{2}^- \bar{\nu}_{\ell_{2}} \,jj$ &\bsmgfoptions{GHTM}\\*
\bf 4106 & $p \overset{\mbox{\tiny{(--)}}}{p} \to H \, jj\to ZZ \, jj\to \ell_{1}^+ \ell_{1}^- \ell_{2}^+ \ell_{2}^- \,jj$ &\bsmgfoptions{GHTM}\\*
\bf 4107 & $p \overset{\mbox{\tiny{(--)}}}{p} \to H \, jj\to ZZ \, jj\to \ell_{1}^+ \ell_{1}^- \nu_{\ell_{2}}  \bar{\nu}_{\ell_{2}} \,jj$ &\bsmgfoptions{GHTM}\\*
&\\*
\hline
&\\*
\bf 4300 & $p \overset{\mbox{\tiny{(--)}}}{p} \to W^{+}W^{-} \to \ell_{1}^{+} \nu_{\ell_{1}} \ell_{2}^{-}\bar{\nu}_{\ell_{2}} $ &\bsmgfoptions{GH}\\*
\bf 4301 & $p \overset{\mbox{\tiny{(--)}}}{p} \to W^{+}W^{-} \to q \bar{q} \, \ell^{-}\bar{\nu}_{\ell} $ &\bsmgfoptions{GHL}\\*
\bf 4302 & $p \overset{\mbox{\tiny{(--)}}}{p} \to W^{+}W^{-} \to \ell^{+} \nu_{\ell} \, q \bar{q} $ &\bsmgfoptions{GHL}\\*
\bf 4330 & $p \overset{\mbox{\tiny{(--)}}}{p} \to ZZ \to \ell_{1}^{-} \ell_{1}^{+}  \ell_{2}^{-} \ell_{2}^{+} $ &\bsmgfoptions{GH}\\*
\bf 4331 & $p \overset{\mbox{\tiny{(--)}}}{p} \to ZZ \to q \bar{q} \, \ell^{-} \ell^{+} $ &\bsmgfoptions{GHL}\\*
\bf 4360 & $p \overset{\mbox{\tiny{(--)}}}{p} \to Z\gamma \to \ell_{1}^{-} \ell_{1}^{+}  \gamma $ &\bsmgfoptions{GH}\\*
\bf 4370 & $p \overset{\mbox{\tiny{(--)}}}{p} \to \gamma\gamma $ &\bsmgfoptions{GH}\\*
&\\* 
\hline
\end{longtable}
}


\begingroup\begin{thebibliography}{10}

\bibitem{Arnold:2008rz}
  K.~Arnold, M.~Bahr, G.~Bozzi {\it et al.},
  ``VBFNLO: A parton level Monte Carlo for processes with electroweak bosons'',
  {\em Comput.\ Phys.\ Commun.}  {\bf 180 } (2009)  1661-1670, \href{http://arxiv.org/abs/0811.4559}{{\tt arXiv:0811.4559}}.

\bibitem{Arnold:2011wj}
  K.~Arnold, J.~Bellm, G.~Bozzi {\it et al.},
  ``VBFNLO: A parton level Monte Carlo for processes with electroweak bosons -- Manual for Version 2.5.0,''
  \href{http://arxiv.org/abs/1107.4038v1}{{\tt arXiv:1107.4038v1}}.
  
\bibitem{Arnold:2011wjv2}
  K.~Arnold, J.~Bellm, G.~Bozzi {\it et al.},
  ``VBFNLO: A parton level Monte Carlo for processes with electroweak bosons -- Manual for Version 2.6.0,''
  \href{http://arxiv.org/abs/1107.4038v2}{{\tt arXiv:1107.4038v2}}.
  
\bibitem{Arnold:2012xn}
  K.~Arnold, J.~Bellm, G.~Bozzi {\it et al.},
  ``Release Note -- Vbfnlo-2.6.0,''
  \href{http://arxiv.org/abs/1207.4975}{{\tt arXiv:1207.4975}}.
  
\bibitem{Campanario:2013qba} 
  F.~Campanario, M.~Kerner, L.~D.~Ninh and D.~Zeppenfeld,
  ``WZ production in association with two jets at NLO in QCD,''
  {\em Phys.\ Rev.\ Lett.} {\bf 111} (2013) 052003,
  \href{http://arXiv.org/abs/1305.1623} {{\tt arXiv:1305.1623}}.

\bibitem{Campanario:2014dpa}
  F.~Campanario, M.~Kerner, L.~D.~Ninh and D.~Zeppenfeld,
  ``Next-to-leading order QCD corrections to $W \gamma$ production in association with two jets,''
  \href{http://arXiv.org/abs/1402.0505} {{\tt arXiv:1402.0505}}.

\bibitem{Campanario:2013gea} 
  F.~Campanario, M.~Kerner, L.~D.~Ninh and D.~Zeppenfeld,
  ``Next-to-leading order QCD corrections to $W^+W^+$ and $W^-W^-$ production in association with two jets,''
  {\em Phys.\ Rev.} {\bf D89} (2014) 054009,
  \href{http://arXiv.org/abs/1311.6738} {{\tt arXiv:1311.6738}}.

\bibitem{kaiser}
N.~Kaiser, `` NLO QCD corrections to $W\gamma$ production via vector boson fusion,'' 
{Diploma Thesis, ITP Karlsruhe 2013}, 
\href{http://www.itp.kit.edu/diplomatheses.en.shtml}{{\tt http://www.itp.kit.edu/diplomatheses.en.shtml}}.
  
\bibitem{Campanario:2013eta}
  F.~Campanario, N.~Kaiser and D.~Zeppenfeld,
  ``$W \gamma$ production in vector boson fusion at NLO in QCD,''
  {\em Phys.\ Rev.} {\bf D89} (2014) 014009,
\href{http://arxiv.org/abs/1309.7259}{{\tt arXiv:1309.7259}}.

\bibitem{Figy:2008zd}
  T.~Figy,
  ``Next-to-leading order QCD corrections to light Higgs Pair production via vector boson fusion,''
  {\em Mod.\ Phys.\ Lett.}\ A {\bf 23} (2008) 1961,
\href{http://www.arXiv.org/abs/0806.2200}{{\tt arXiv:0806.2200}}.

\bibitem{Baglio:2012np}
  J.~Baglio, A.~Djouadi, R.~Groeber, M.~M.~Muehlleitner, J.~Quevillon and M.~Spira,
  ``The measurement of the Higgs self-coupling at the LHC: theoretical status,''
  {\em JHEP} {\bf 1304} (2013) 151,
\href{http://www.arXiv.org/abs/1212.5581}{{\tt arXiv:1212.5581}}.

\bibitem{robin}
R.~Roth, ``NLO QCD corrections to $WH$ + jet production at the LHC,'' 
{Diploma Thesis, ITP Karlsruhe 2013}, 
\href{http://www.itp.kit.edu/diplomatheses.en.shtml}{{\tt http://www.itp.kit.edu/diplomatheses.en.shtml}}.
  
\bibitem{semilep}
B.~Feigl, ``Electroweak Processes in the Standard Model and Beyond: Backgrounds to Higgs Physics and Semileptonic Decay Modes '', {PhD Thesis, ITP Karlsruhe 2013}, 
\href{http://digbib.ubka.uni-karlsruhe.de/volltexte/1000037298}{{\tt http://digbib.ubka.uni-karlsruhe.de/volltexte/1000037298}}.

\bibitem{Beringer:1900zz}
  J.~Beringer {\it et al.}  [Particle Data Group Collaboration],
  ``Review of Particle Physics (RPP),''
   {\em Phys.\ Rev.} {\bf D86} (2012) 010001.

\bibitem{Frank:2012wh}
  J.~Frank, M.~Rauch and D.~Zeppenfeld,
  ``Spin-2 Resonances in Vector-Boson-Fusion Processes at NLO QCD'',
  Phys.\ Rev.\ {\bf D87}, 055020 (2013)
 \href{http://arxiv.org/abs/1211.3658}{{\tt arXiv:1211.3658}}. 
  
\bibitem{Frank:2013gca}
  J.~Frank, M.~Rauch and D.~Zeppenfeld,
  ``Higgs Spin Determination in the WW channel and beyond''
 \href{http://arxiv.org/abs/1305.3658}{{\tt arXiv:1305.3658}}. 

\bibitem{Nhung:2013jta}
  D.~T.~Nhung, L.~D.~Ninh and M.~M.~Weber,
  ``NLO corrections to WWZ production at the LHC,''
  {\em JHEP } {\bf 1312} (2013) 096,
\href{http://www.arXiv.org/abs/1307.7403}{{\tt arXiv:1307.7403}}.

\end{thebibliography}\endgroup
\end{document}